\documentclass[]{aa}
\usepackage{amsmath}
\usepackage{amssymb}
\usepackage{natbib}
\usepackage{graphicx}
\usepackage{txfonts}
\usepackage[english]{babel}
\usepackage{hyperref}
\usepackage{units}
\hypersetup{colorlinks=true, linkcolor=blue, citecolor=blue, urlcolor=blue}
\usepackage{color}

\begin{document}

\title{Multiwavelength optical properties of compact dust aggregates in protoplanetary disks}
\author{
        M. Min\inst{1,2}
                \and
        Ch. Rab\inst{3}
                \and
        P. Woitke\inst{4}
                \and
        C. Dominik\inst{2}
                \and
        F. M\'enard\inst{5}
}

\institute{
SRON Netherlands Institute for Space Research, Sorbonnelaan 2, 3584 CA Utrecht, The Netherlands
        \and
Astronomical institute Anton Pannekoek, University of Amsterdam,
Science Park 904, 1098 XH, Amsterdam, The Netherlands
        \and
University of Vienna, Dept. of Astrophysics, T\"urkenschanzstr. 17, 1180
Wien, Austria
        \and
SUPA, School of Physics \& Astronomy, University of St. Andrews, North Haugh, St. Andrews KY16 9SS, UK
        \and
UMI-FCA, CNRS/INSU, France (UMI 3386), and Dept. de Astronom\'{\i}a, Universidad de Chile, Santiago, Chile
}

   \date{\today}

\offprints{M. Min, \email{\href{mailto:M.Min@uva.nl}{M.Min@uva.nl}}}

  \abstract
   {
   In protoplanetary disks micron-size dust grains coagulate to form larger structures with complex shapes and compositions. 
   The coagulation process changes the absorption and scattering properties of particles in the disk in significant ways.
   To properly interpret observations of protoplanetary disks and to place these observations in the context of the first steps of planet formation, it is crucial to understand the optical properties of these complex structures.
   }
   {
   We derive the optical properties of dust aggregates using detailed computations of aggregate structures and compare these computationally demanding results with \textit{\textup{approximate}} methods that are cheaper to compute in practice.  In this way we wish to understand the merits and problems of approximate methods and define the context in which they can or cannot be used to analyze observations of objects where significant grain growth is taking place.
   }
   {For the detailed computations we used the discrete dipole approximation (DDA), a method able to compute the interaction of light with a complexly shaped, inhomogeneous particle. 
   We compared the results to those obtained using spherical and irregular, homogeneous and inhomogeneous particles.
   }
   {
    While no approximate method properly reproduces all characteristics of large dust aggregates, the thermal properties of dust can be analyzed using irregularly shaped, porous, inhomogeneous grains.  The asymmetry of the scattering phase function is a good indicator of aggregate size, while the degree of polarization is probably determined by the size of the constituent particles.  Optical properties derived from aggregates significantly differ from the most frequently used standard (``astronomical silicate'' in spherical grains).
   We outline a computationally fast and relatively accurate method that can be used for a multiwavelength analysis of aggregate dust in protoplanetary disks.
   }
{}

   \keywords{scattering -- techniques: polarimetric -- protoplanetary disks -- circumstellar matter}

\maketitle

\section{Introduction}

In high-density environments, such as protoplanetary disks, dust grains stick together to form larger structures. This coagulation process is the starting point of the formation of larger bodies. The growth of dust grains to larger aggregates affects their dynamical \citep[e.g.,][]{2007A&A...461..215O} and optical \citep[e.g.,][]{1996A&A...311..291H, 2006A&A...445.1005M, 2007A&A...470..377V, 2014A&A...568A..42K} properties. The opacity of the disk is dominated almost everywhere by the dust particles. Therefore, with the changing absorption efficiency with coagulation state, the thermal structure of the disk and the internal radiation field also change. This in turn influences the chemistry and the dynamical properties.  Therefore, the optical properties of the dust particles in a protoplanetary disk form the basis of any modeling or interpretation attempt. Here we make a step toward a better understanding and providing a computationally feasible tool to compute realistic optical properties of aggregated particles.

There is extensive literature on the optical properties of complex dust particles (see, e.g., the excellent books on this topic by \cite{1983asls.book.....B, 1957lssp.book.....V} or, e.g., \citet{1988ApJ...333..848D, 2007ApOpt..46.4065V, 2009ASPC..414..356M} and other papers by these authors). Much is known about general properties, trends and characteristics of the optics of these particles as a function of wavelength, particle shape, composition, and size. The large body of knowledge on this is in sharp contrast with what is usually applied in radiative transfer modeling of dusty astronomical environments. The underlying reason is that exact computations of the optical properties of complex particles is computationally very demanding.  Here we aim at bridging the gap between fundamental knowledge and practical application of the optical properties of complex dust aggregates by analyzing the properties of exact computations and providing a computationally feasible method to reproduce these properties for inclusion in radiative transfer modeling.

The optical properties of particles are often computed using the Mie theory \citep{1908AnP...330..377M}. With this approach the particles are assumed to be homogeneous spherical particles, which usually is an inaccurate approximation for the optical properties of complexly shaped particles \citep[see, e.g.,][]{2005A&A...432..909M}. The applicability and accuracy of the Mie theory can be somewhat increased by combining it with effective medium theory to  mix different materials together or include porosity \citep[see, e.g.,][]{2005A&A...429..371V}. For aggregates with an extremely open (fluffy) structure, effective medium theory provides a reasonable approximation. \citet{2014A&A...568A..42K} gave a recipe for approximating the optical properties of such extreme aggregates. An improvement to the effective medium approach, simulating aggregates with irregularly shaped constituents, can be performed by combining their recipe with the aggregate polarizability mixing rule (APMR) from \cite{2008A&A...489..135M}.

Exact methods for computing the optical properties of aggregated particles do exist. However, the aggregate size that can be handled by  these methods is limited because these methods are very demanding in terms of cpu and memory requirements. Methods like the superposition T-matrix \citep{1996JOSAA..13.2266M} or the discrete dipole approximation \citep[DDA;][]{1973ApJ...186..705P, 1994JOSAA..11.1491D} are used to gain insight into the optical properties of aggregates and how they differ from those of compact particles. This is usually done with a specific wavelength range or observational characteristic in mind. Within the framework of the FP7 project DIANA we aim at modeling protoplanetary disks using a broad wavelength range and a wide variety of observation types (dust and gas diagnostics) at different spatial resolution. This requires a method for the optical properties of the dust particles that is able to produce accurate results for all characteristics.

In this paper we study aggregate particles over a wide range of wavelengths, using absorption, scattering, and polarization characteristics.  By combining these various optical properties we aim to provide combinations of observational characteristics that are unique for aggregate particles. In addition, we study which approximate computational tools catch the optical properties of aggregate particles best and make recommendations for modeling aggregate particles in protoplanetary disks.

\section{Optical properties}
\label{sec:optical properties}

For convenience we first introduce a few definitions and terms that are used throughout the paper.

The first important characteristic of a particle is its size. For a spherical particle this is easily described by its radius, $r$. For a nonspherical particle we consider the size of the particle to be the radius of a homogeneous sphere with the same material volume (i.e., the same mass), denoted by $r_V$. Thus, $r_V$ is defined as $r_V=\sqrt[3]{3V/(4\pi)}$, with $V$ the total material volume (i.e., the same mass) of the particle.

The optical properties of a particle can be defined in terms of the scattering matrix and the cross sections for absorption and scattering. The scattering matrix is a $4\times4$ matrix that converts the Stokes vector of the incoming light into the Stokes vector of the outgoing light. The angle-dependent elements of the scattering matrix are denoted by $F_{ij}(\theta)$, where $\theta$ is the scattering angle. This angle is defined such that $\theta=0^\circ$ refers to forward scattering and $\theta=180^\circ$ refers to backward scattering. The $F_{11}$ element of the scattering matrix is commonly referred to as the phase function. This element determines in which direction the energy is redistributed. The ratio $-F_{12}/F_{11}$ represents the degree of linear polarization for unpolarized incoming light. For the cross sections we use the symbol $\kappa_\mathrm{abs}, \kappa_\mathrm{sca}, \kappa_\mathrm{ext}$ representing the cross section per unit mass for absorption, scattering, or extinction.

\begin{figure*}[!tp]
\centerline{\resizebox{0.9\hsize}{!}{\includegraphics{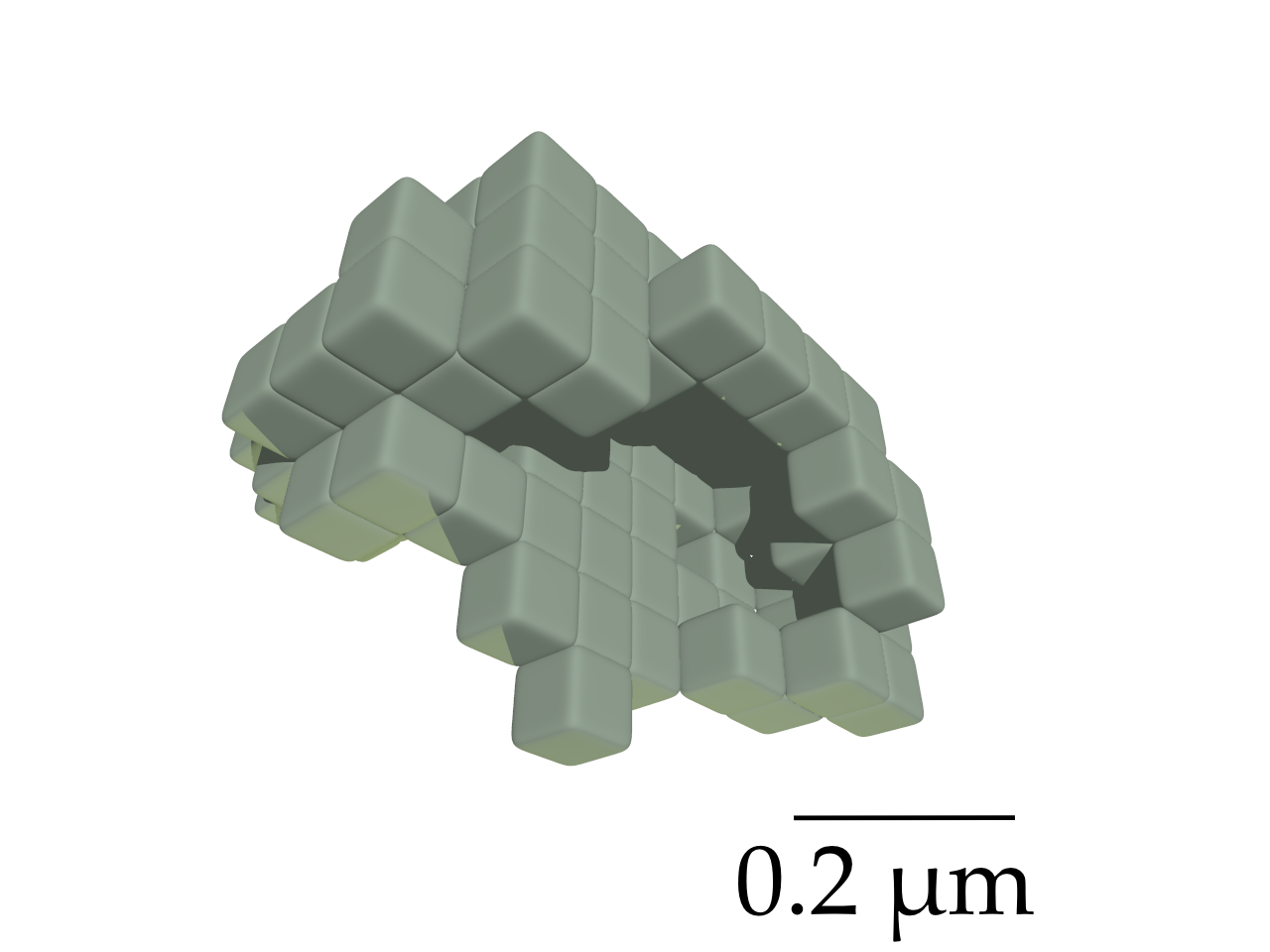}\includegraphics{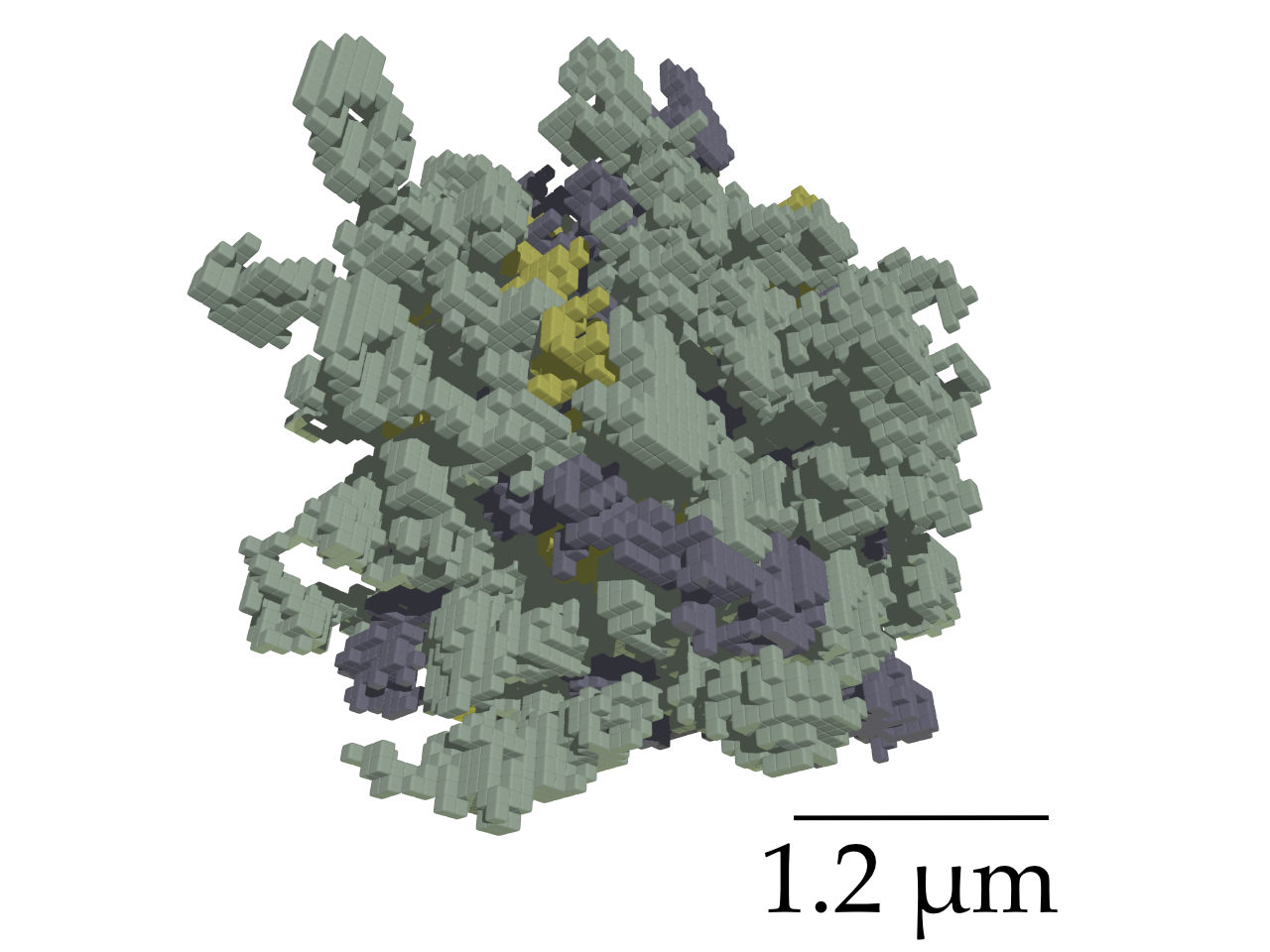}\includegraphics{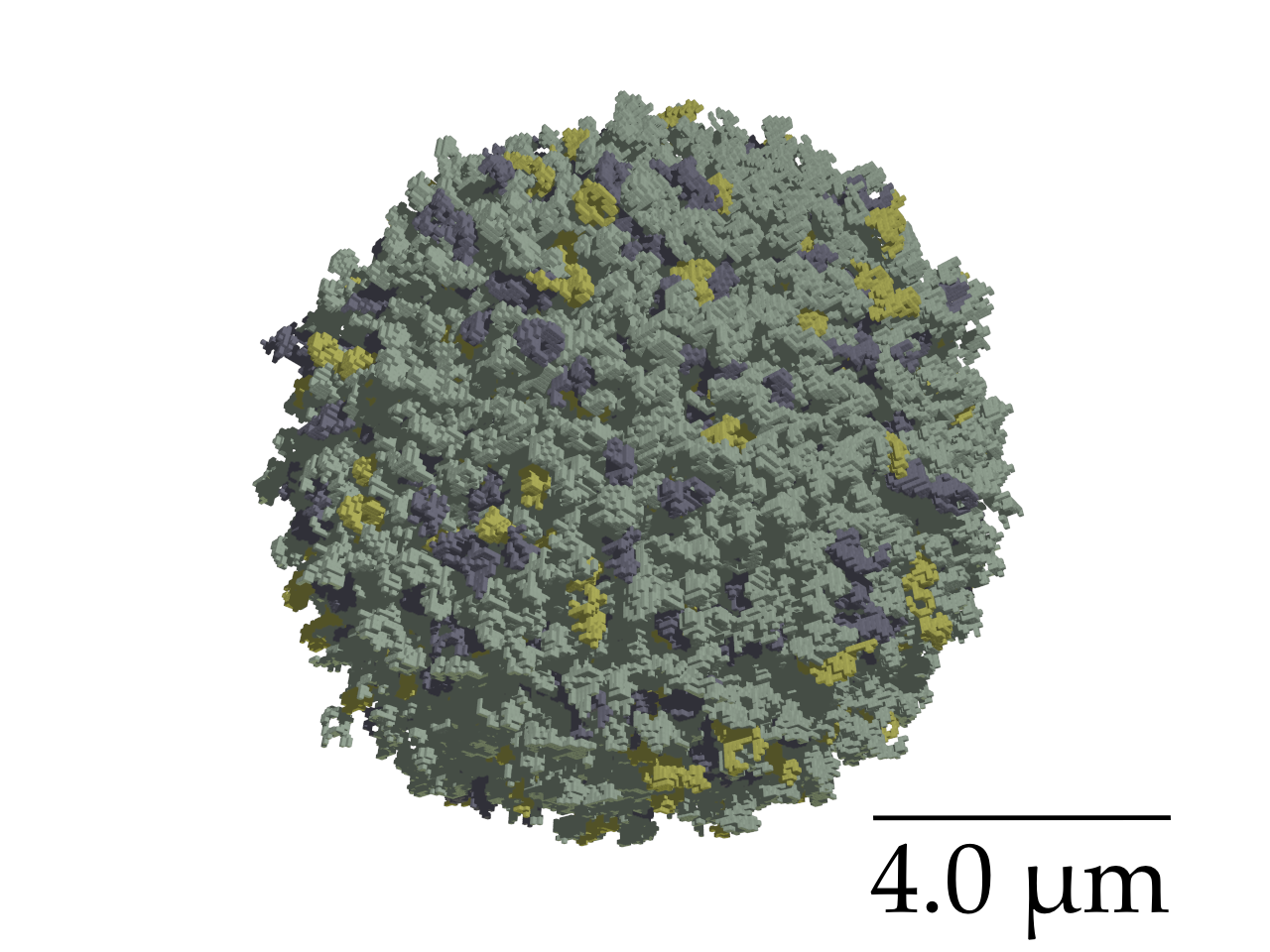}}}
\caption{Images of the particles used in the DDA computations. The left image shows a single monomer, containing approximately 100 dipoles, the middle image is an aggregate containing 216 of these monomers, and at the far right we show the largest aggregate we consider, containing 8000 monomers. The horizontal line indicates the scale in each image.}
\label{fig:aggregates}
\end{figure*}

\subsection{Exact method: Discrete dipole approximation}

To computate the optical properties of the aggregates we used
the DDA \citep[][]{1973ApJ...186..705P, 1994JOSAA..11.1491D}. This is a computational method that allows obtaining the optical properties of arbitrarily shaped particles with an arbitrary compositional mix. In the DDA the particle volume is represented as a collection of dipoles. Each dipole represents a subvolume element of the particle, the discretization of the integral over the volume of the particle. The interaction of these dipoles with each other is used to compute the interaction of the different volume elements in the particle. For a recent, detailed overview of the DDA we refer to \citet{2007JQSRT.106..558Y}. Despite the somewhat confusing term approximation in the name, the method gives an exact solution to the Maxwell equations in the limit of an infinitely large number of dipoles \citep{1992ApJ...394..494L, Lakhtakia1993}. Therefore, the method itself is exact, but the numerical implementation provides an approximation.  Especially for high values of the refractive index, where the method is still exact in principle, the number of dipoles has to be large enough to trace the large gradients of the electromagnetic field inside the particle.
For the computations we used the publicly available code \textsc{ADDA} \citep{2011JQSRT.112.2234Y}. This code has the advantage that it is very fast, can be run in parallel mode, and has improved accuracy for large refractive indices \citep{2010PhRvE..82c6703Y}.

The accuracy of DDA depends on the number of subvolume elements (dipoles) chosen to represent the particle. First of all, it is important to have enough dipoles to represent the shape of the particle that is to be modeled. Second, it is important that the dipoles are small enough, such that the electromagnetic field inside a single dipole can be considered constant. This means that the dipoles need to be significantly smaller than the wavelength of radiation \textit{\textup{both inside and outside}} of the particle. Since DDA basically solves a $3N$x$3N$ matrix equation, the number of  dipoles is given by the available computing power, which limits the size of particles for which computation is feasible. Modern DDA programs use a fast Fourier transform method to speed up the matrix solution. This approach allows handling large particles, with the drawback that the dipoles have to be placed on a regular, rectangular grid.  This makes computations of very fluffy aggregates (with void dipoles) much more challenging than very compact particles of the same mass. 

In reality aggregates are build out of different monomers with different sizes, shapes, and compositions. In our DDA computations we used a single size for all monomers, but we gave each monomer a different shape by sampling each aggregate monomer with a number of dipoles.

For the smaller aggregates we considered several different aggregate configurations, while for the largest aggregates ($r_V>3\,\mu$m) we only performed the computations for a single aggregate. The optical properties were always computed for a random set of orientations, determined by the random orientation sampling in the \textsc{ADDA} code. This limited set of aggregate realizations was chosen to make the computations feasibile, but it can cause residual noise on the computed polarization and intensity phase curves. However, this does not hamper the interpretation of general trends.

\subsubsection{Creating the aggregates}

To compute realistic optical properties of cosmic dust aggregates, we have to start by creating the aggregates. Many computations on dust aggregates have been performed using spherical monomers. However, \citet{2008A&A...489..135M} have shown that the shapes of the monomers of the aggregate are very important for the optical properties of the aggregate as a whole. Therefore, we here used irregularly shaped monomers.  We used Gaussian random field particles \citep[GRF; see][]{2007A&A...462..667M} to construct monomers using about 100 subvolume elements (dipoles).  One hundred dipoles allow for a reasonably irregular shape while limiting the number of dipoles per monomer. We constructed 100 realizations of these GRF particles, all with different randomly determined shapes.  The aggregates were then constructed as follows.
\begin{enumerate}
\item We started with a single monomer, randomly chosen from the 100 realizations of GRF particles.
\item We placed another monomer (also randomly chosen) at the center of the first so that their volumes overlapped.
\item The second monomer was moved by random brownian motion until the volume no longer overlapped. The monomers were allowed
to have several contacting dipoles, but no overlap.
\item Steps 2 and 3 were repeated for each monomer that was to
be added to the aggregate.
\end{enumerate}

This procedure creates relatively compact aggregates (see also Fig.~\ref{fig:aggregates}), allowing efficient use of DDA. The interactions between the monomers in the aggregate are also stronger for more compact aggregates, making the effects of aggregation more visible in the optical properties. 

Each monomer in the aggregate is made of a single material. We randomly assigned a material to each monomer using the overall composition of 75\% silicate, 15\% carbon, and 10\% iron sulfide (by volume). This composition is roughly consistent with the solar system composition proposed by \citet{2011Icar..212..416M}. We used the refractive index data from \citet{1995A&A...300..503D}, \citet{1993A&A...279..577P}, and \citet{1994ApJ...423L..71B} for the silicate (MgSiO$_3$), carbon, and iron sulfide particles, respectively.

Since both amorphous carbon and iron sulfide have very large refractive indices, we switched to the filtered coupled dipole mode of ADDA to increase the efficiency and accuracy \citep{2010PhRvE..82c6703Y}. The largest particle we considered was an aggregate with a volume-equivalent radius of $r_V=4\,\mu$m composed of 8000 monomers with $r_V=0.2\,\mu$m. As each monomer is built from 100 dipoles, we have 800.000 occupied dipoles in the DDA grid.

We assumed the aggregates in protoplanetary disks to be oriented randomly, therefore we computed for each aggregate realization the orientation-averaged optical properties.

\subsection{Approximate methods}

The exact DDA computations of aggregates provide detailed properties of aggregate particles. However, for day-to-day use in radiative transfer modeling, for instance, these computations are too slow. Therefore, approximate methods that capture the essence of the exact computations are usually relied upon. For the approximate methods in this paper we considered two possibilities for the external shape and two for the internal structure and material mixing. This gave us four different methods of different complexity. All these methods are computationally much less demanding than the DDA method described above; the multiwavelength optical properties of a single particle can always be computed within a minute.

\subsubsection{Particle shape}

For the shape of the particles we used two representations: 1) homogeneous spheres, and 2) very irregularly shaped particles. For the homogeneous spheres we used the Mie theory to compute the optical properties \citep{1908AnP...330..377M}. The optical properties of an irregular particle were approximated by averaging over a distribution of hollow spheres \citep[DHS;][]{2005A&A...432..909M}. While the DHS method considers hollow spherical particles, it has been shown to be very effective in reproducing the properties of natural samples of complex particle shapes, including the correct positions of features related to solid-state resonances \citep[see, e.g.,][]{2001A&A...378..228F, 2003A&A...404...35M, 2009A&A...504..875M}.
For the DHS particles we used an irregularity parameter, $f_\mathrm{max}=0.8$. Computationally, $f_\mathrm{max}$ represents the maximum volume fraction occupied by the central void in the hollow sphere. In practice, this is a parameter representing the more general amount of deviation from a perfect homogeneous sphere.

\begin{figure*}[!t]
\centerline{\resizebox{0.85\hsize}{!}{\includegraphics{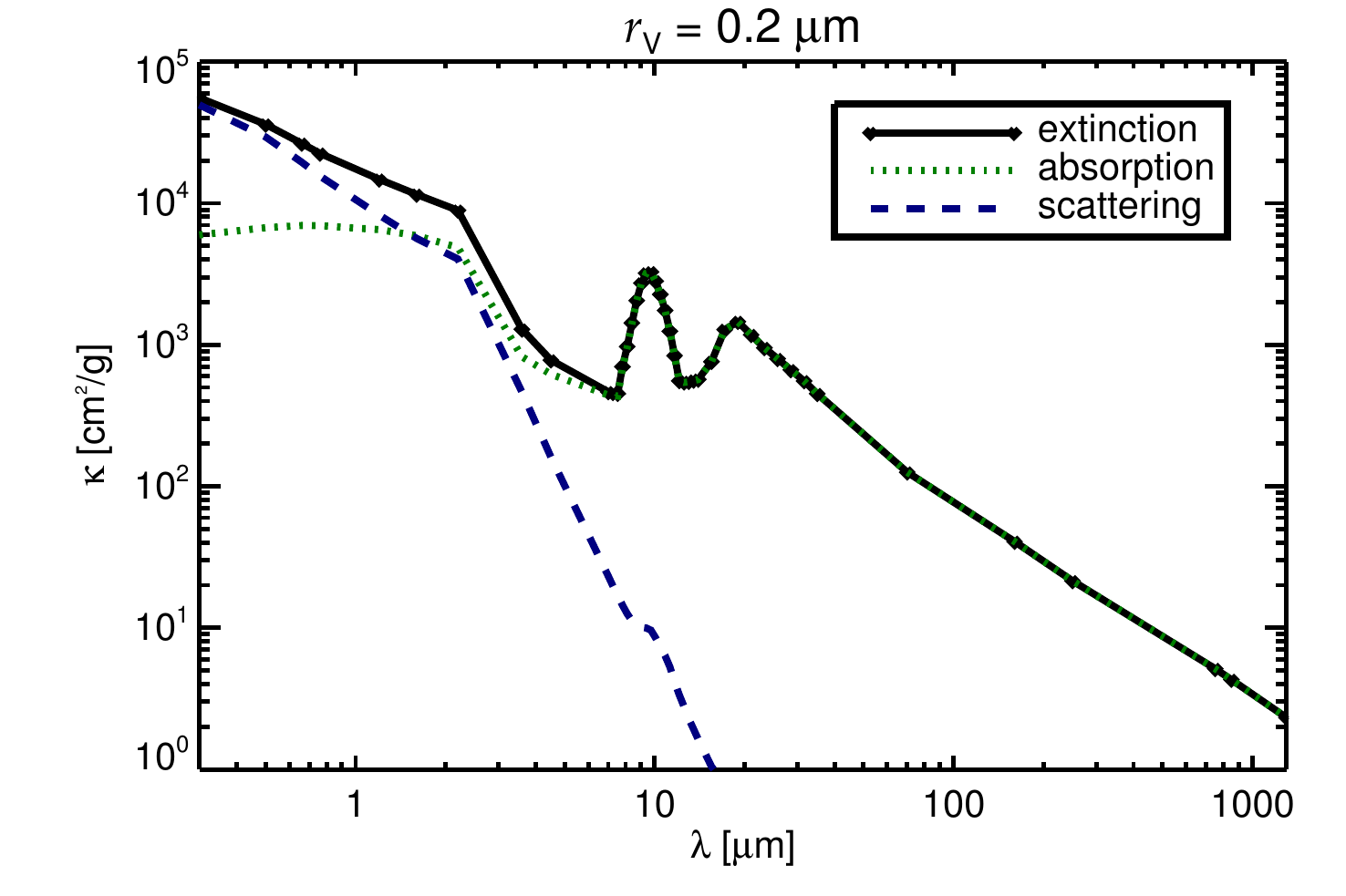}\includegraphics{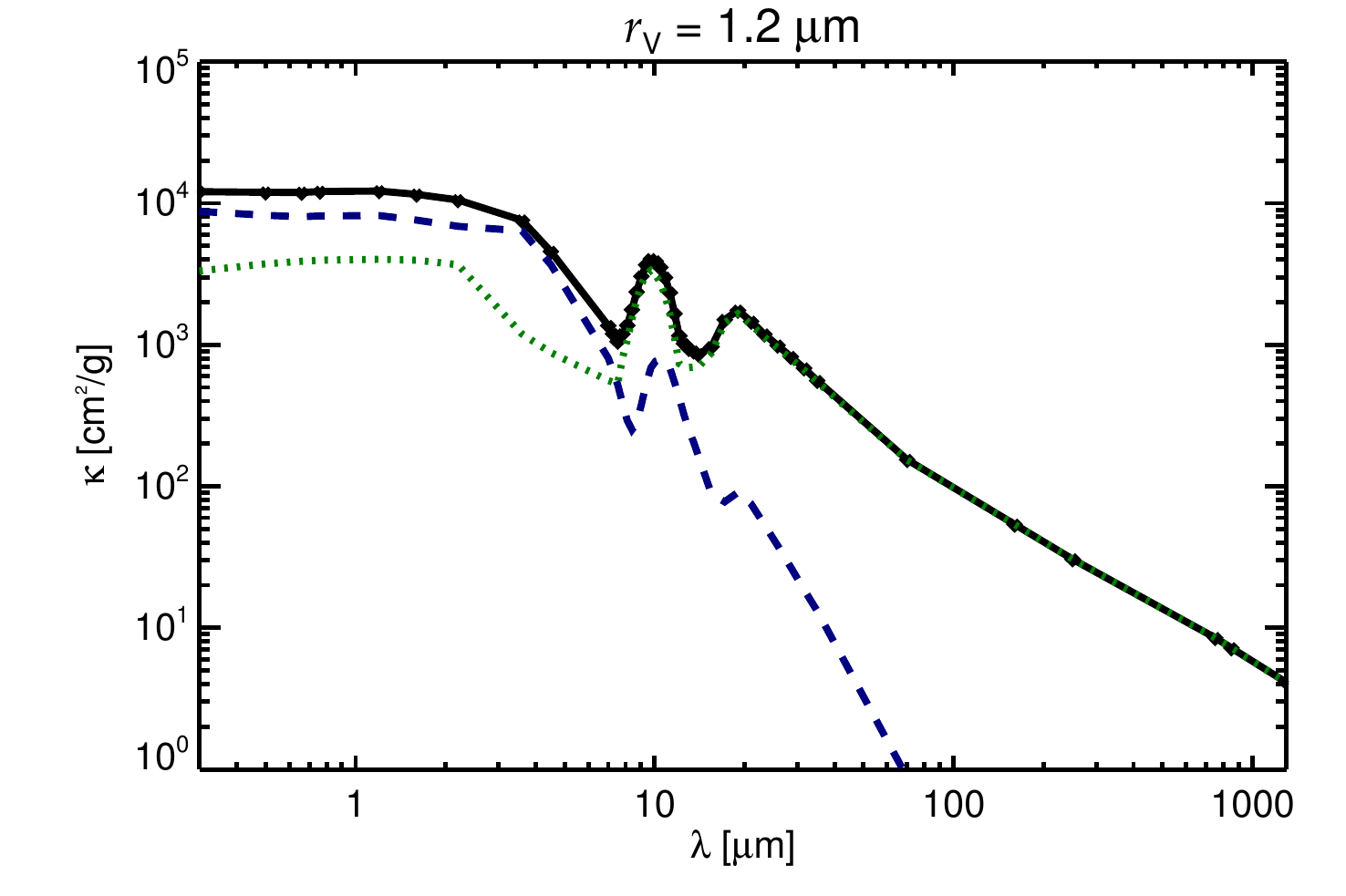}}}
\centerline{\resizebox{0.85\hsize}{!}{\includegraphics{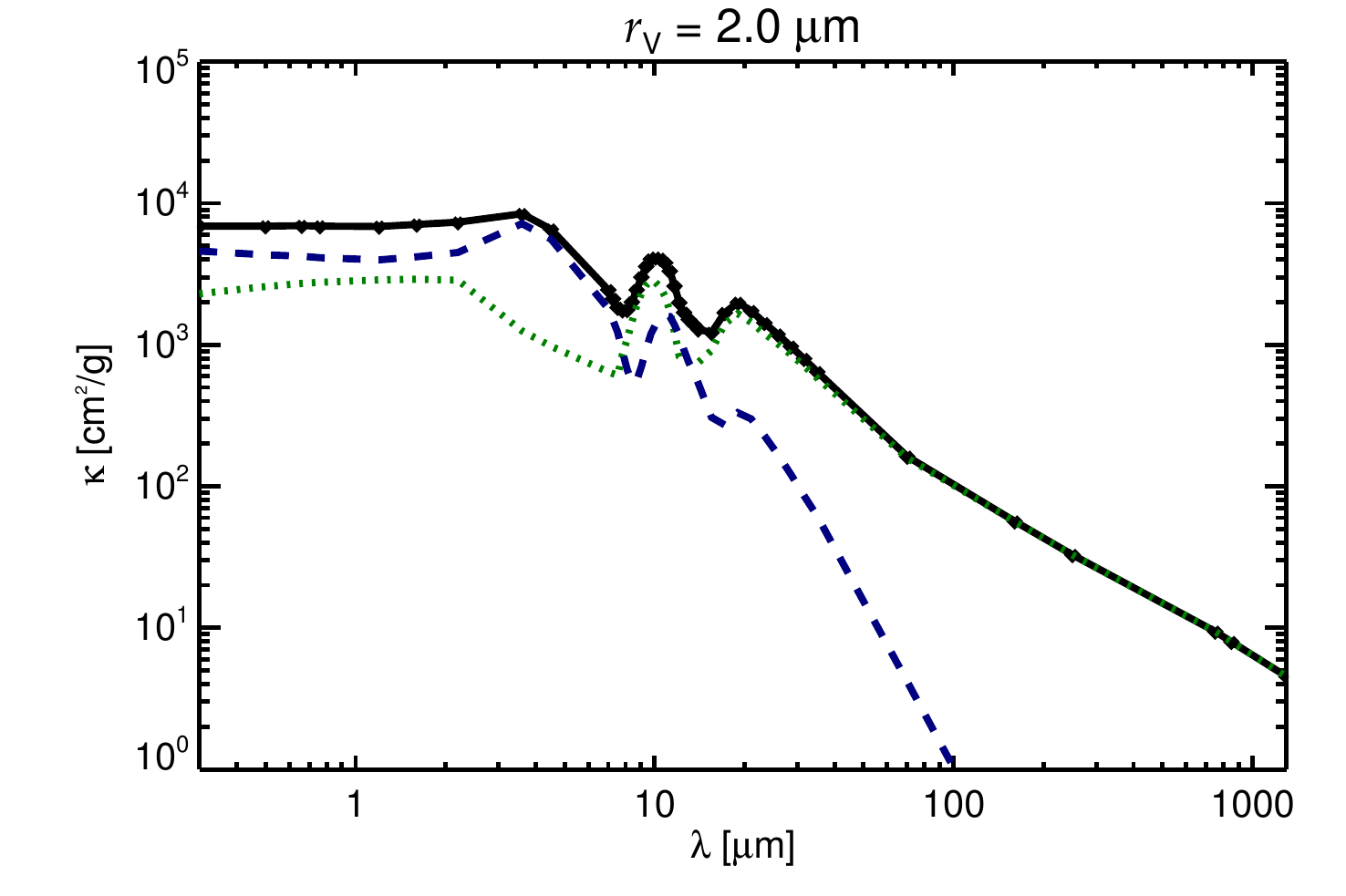}\includegraphics{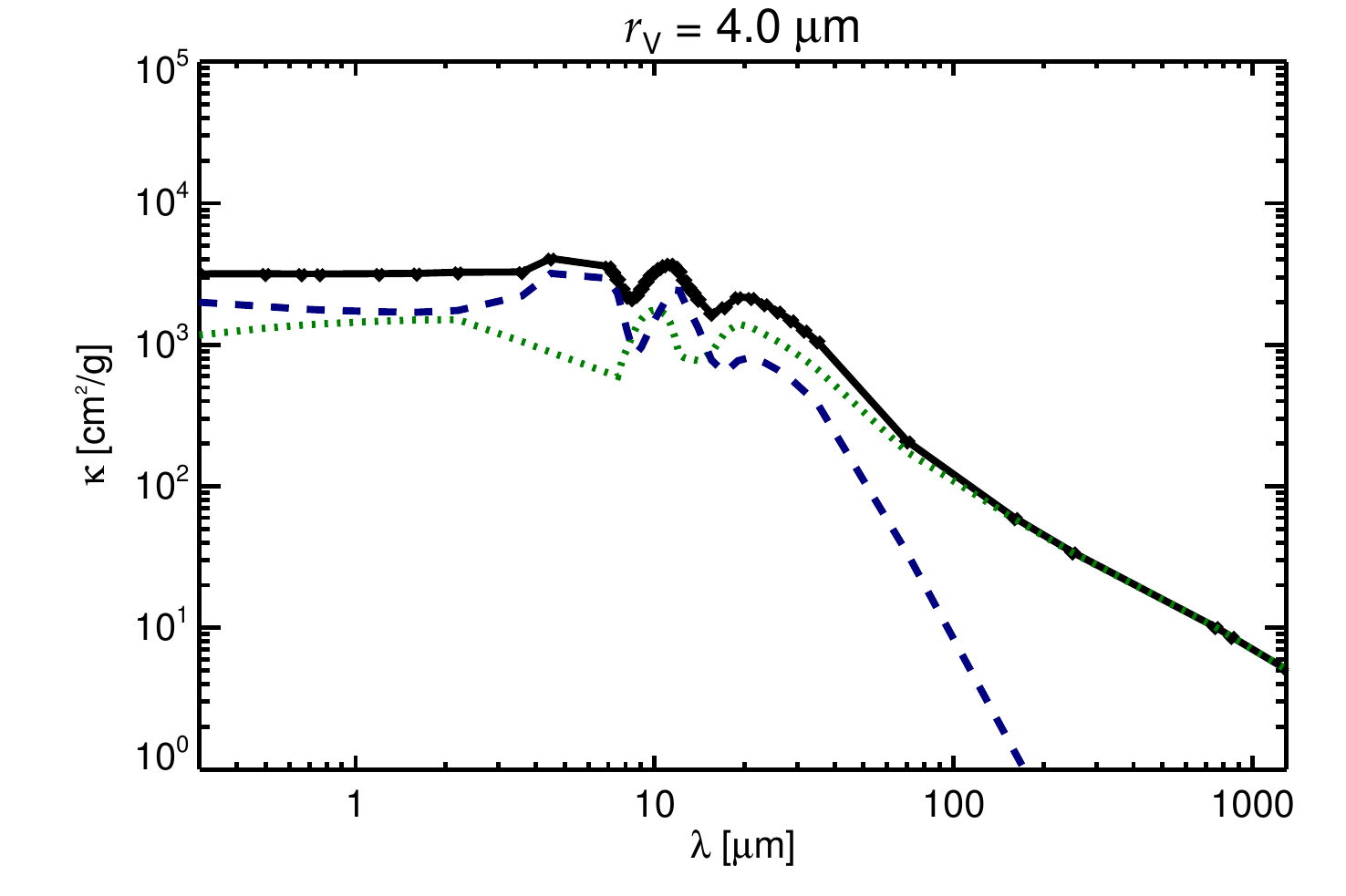}}}
\caption{Cross sections per unit mass of the aggregate particles computed using DDA. We plot the cross sections for extinction, absorption, and scattering. The upper left panel shows an effective size of $r_\mathrm{V}=0.2\,\mu$m (single grains), the upper right panel $r_\mathrm{V}=1.2\,\mu$m, the lower left panel $r_\mathrm{V}=2.0\,\mu$m, and the lower right panel $r_\mathrm{V}=4.0\,\mu$m.}
\label{fig:crosssections}
\end{figure*}

\subsubsection{Internal structure and material mixing}

In a realistic aggregate each monomer is composed of roughly one material, and these components together make up the aggregate. This mixture can be considered in two extreme ways: either the homogeneous monomers dominate, or the mixed-composition aggregate dominates. 
These two extremes can be captured with two different approaches that are commonly used for multi-composition mixtures: 1) adding the opacities of homogeneous grains, and 2) effective medium computations of mixtures at the smallest scale. Method 1 is often used in analyses of the mineralogy of disks, where the composition of the monomers is considered that make up the aggregates. In this method a mix of materials for each component is assumed, sometimes with different size distributions. Here, for simplicity, we assumed that the size distribution of the different materials is the same. Method 2, on the other hand, is often used in radiative transfer modeling where the overall, average absorption and reemission efficiencies of the aggregates as a whole are to be considered. Here we compare the two methods using the optical properties computed with DDA. In effective medium computations the refractive index of the particle is replaced with an effective refractive index computed from the refractive indices of the submaterials and their abundances. We used the Bruggeman mixing rule. Like most effective medium theories, it assumes that the mixing takes place at the finest level, such that it is impossible to distinguish different material components.   Effective medium theories   easily allow including microporosity by mixing in vacuum. We added 25\% of vacuum to the particles computed with the effective medium theory. For the computations in method 1 we did not consider any porosity, that is, the particles are solid and homogeneous.

\section{Results}
\label{sec:results}

The DDA aggregate computations were performed on a parallel computer. The largest aggregate, 8000 monomers, has 800.000 occupied grid cells. To improve the accuracy of the computations, we split each of these grid cells using the refinement method included in ADDA until convergence in the optical properties was reached at a level of $10^{-5}$. The computation of the largest aggregate took four days of CPU time using 64 cores for 44 wavelength points.

\begin{figure*}[!t]
\centerline{\resizebox{0.85\hsize}{!}{\includegraphics{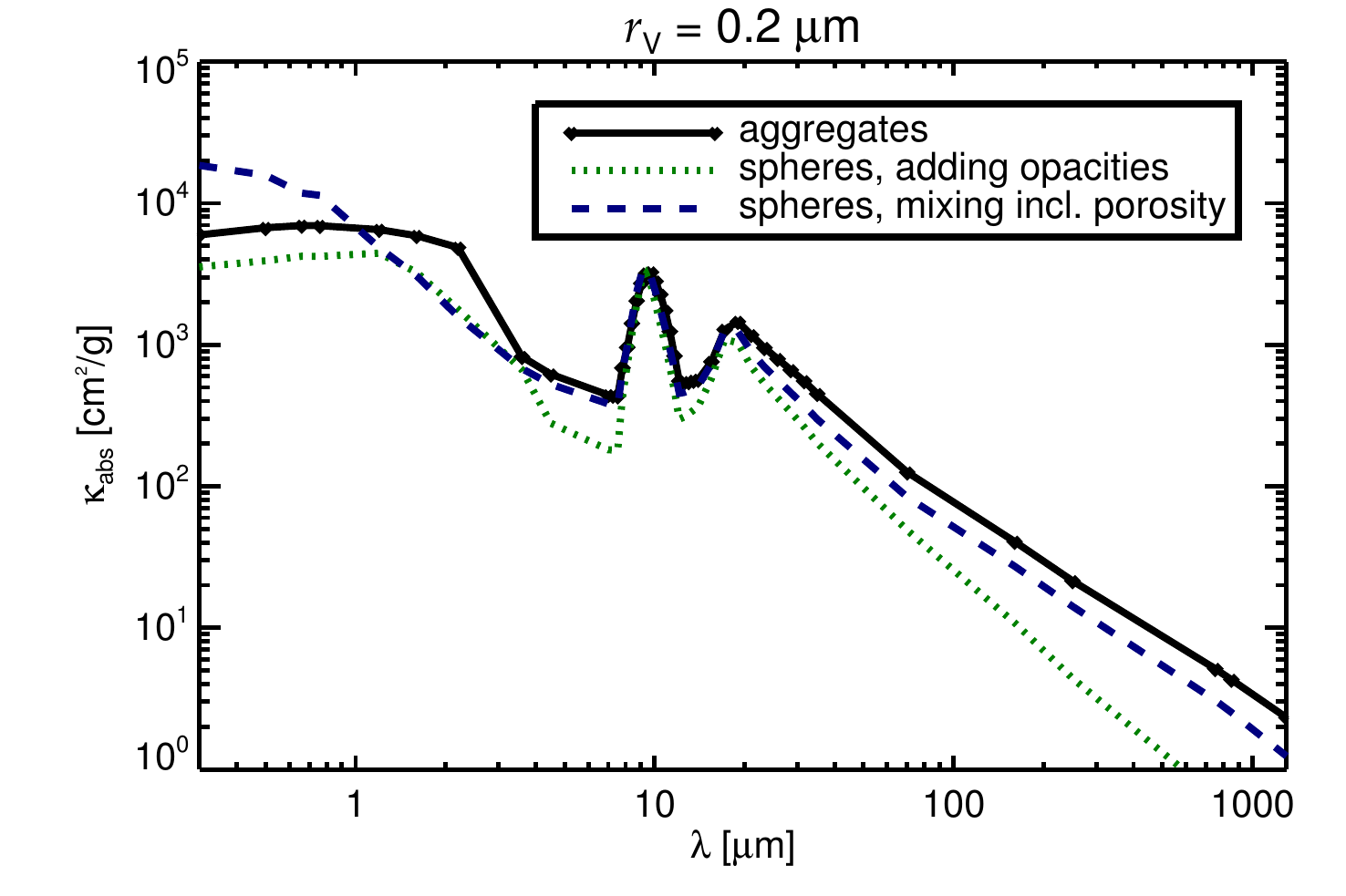}\includegraphics{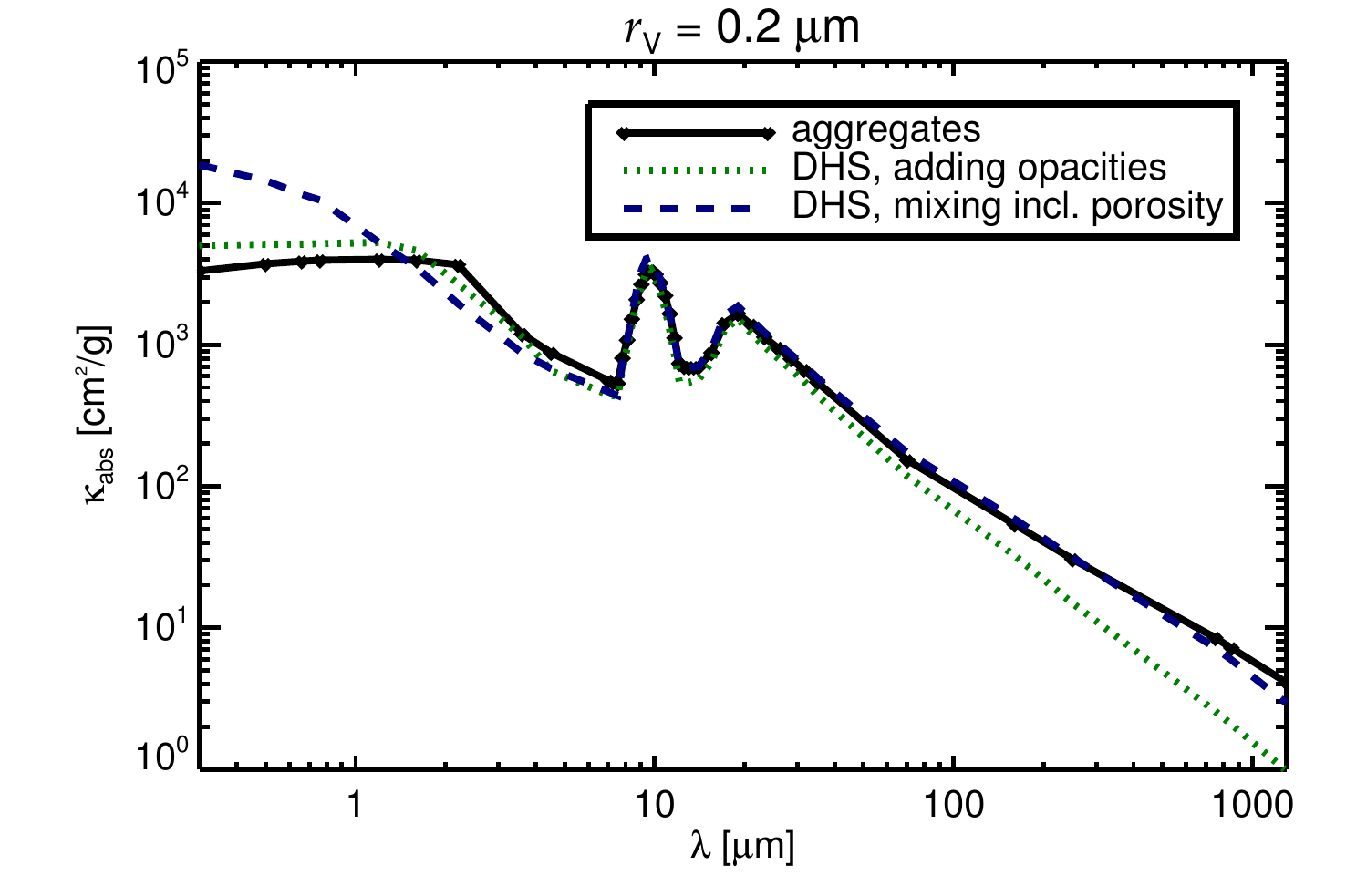}}}
\centerline{\resizebox{0.85\hsize}{!}{\includegraphics{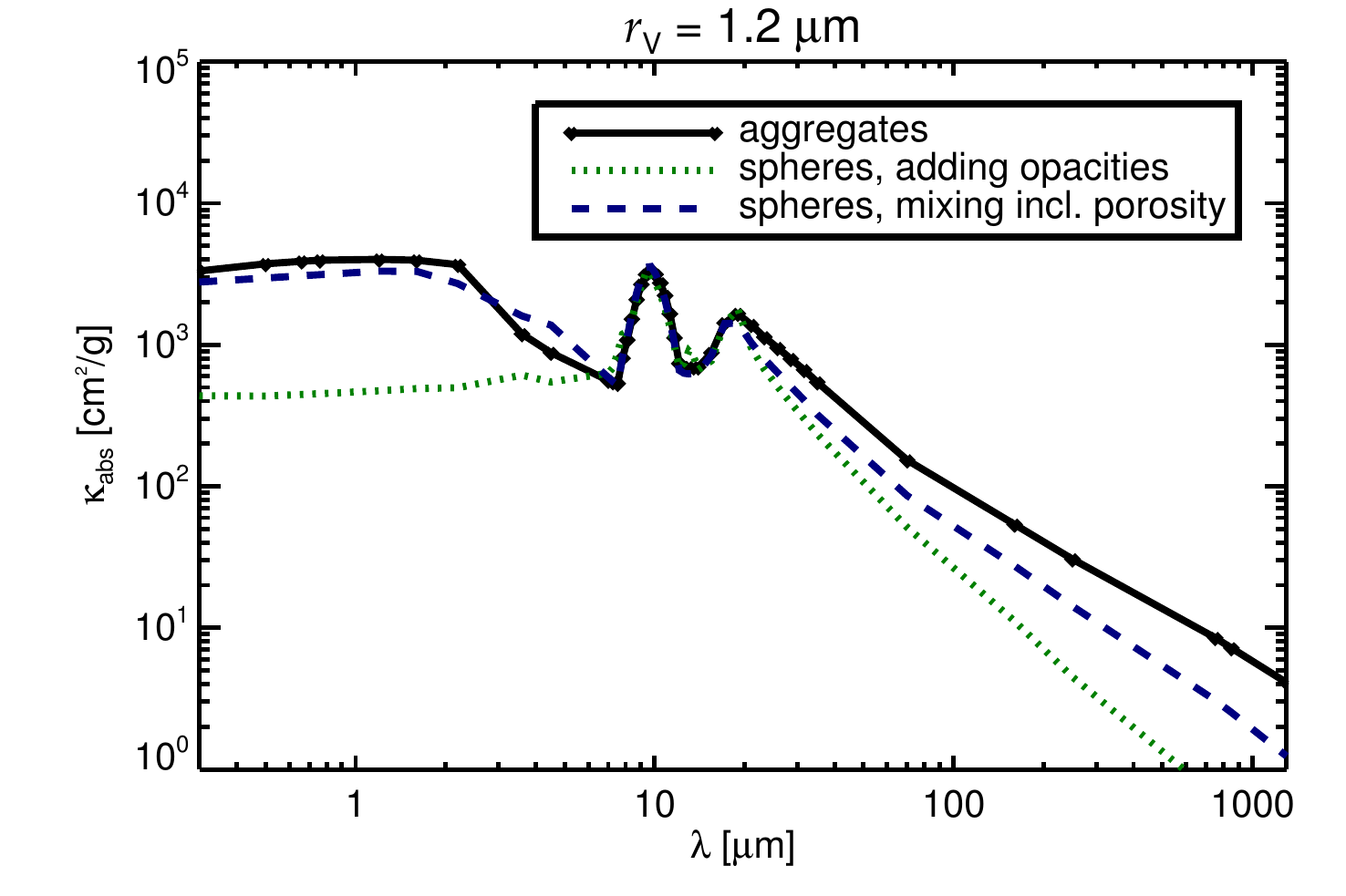}\includegraphics{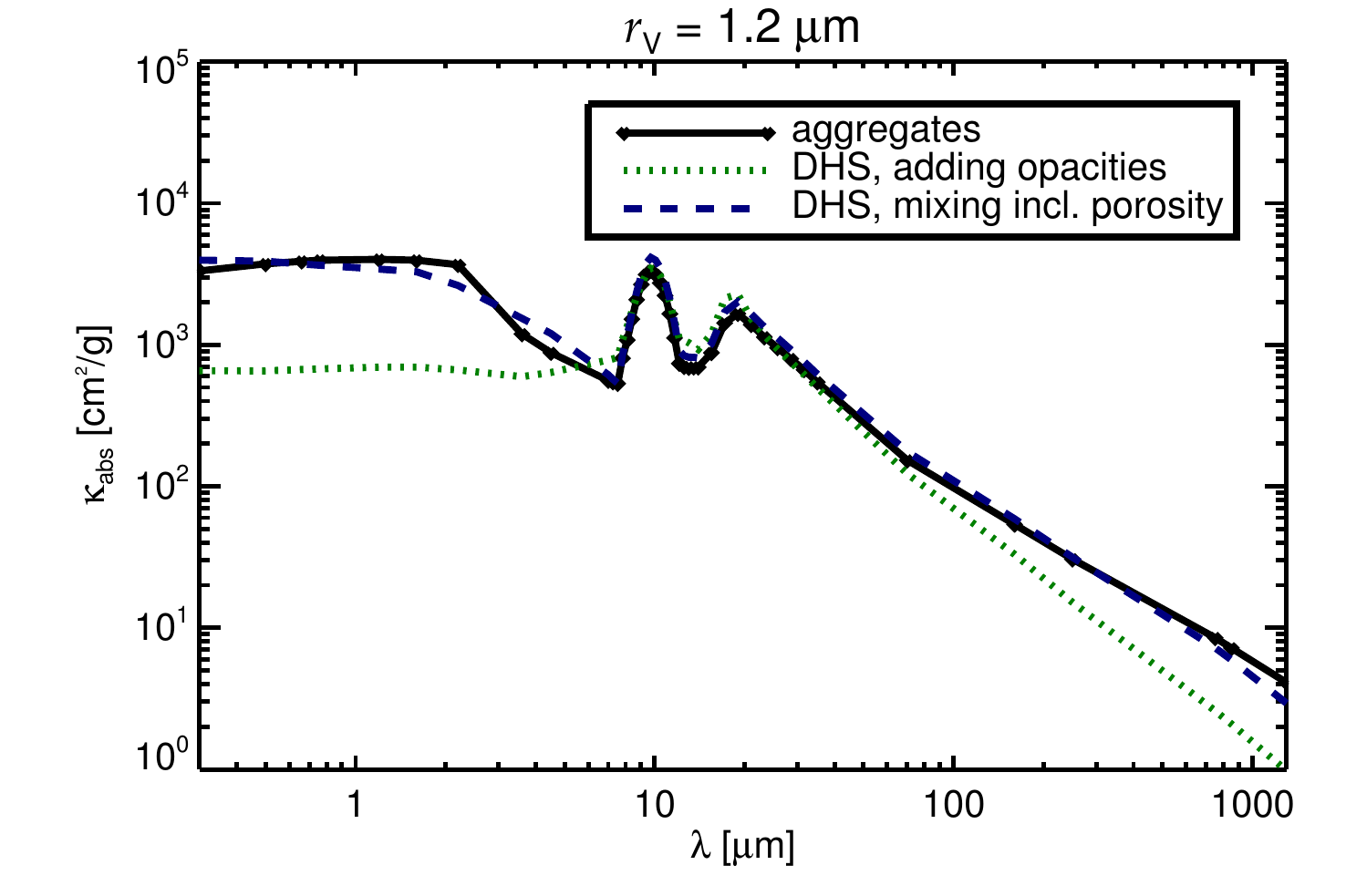}}}
\centerline{\resizebox{0.85\hsize}{!}{\includegraphics{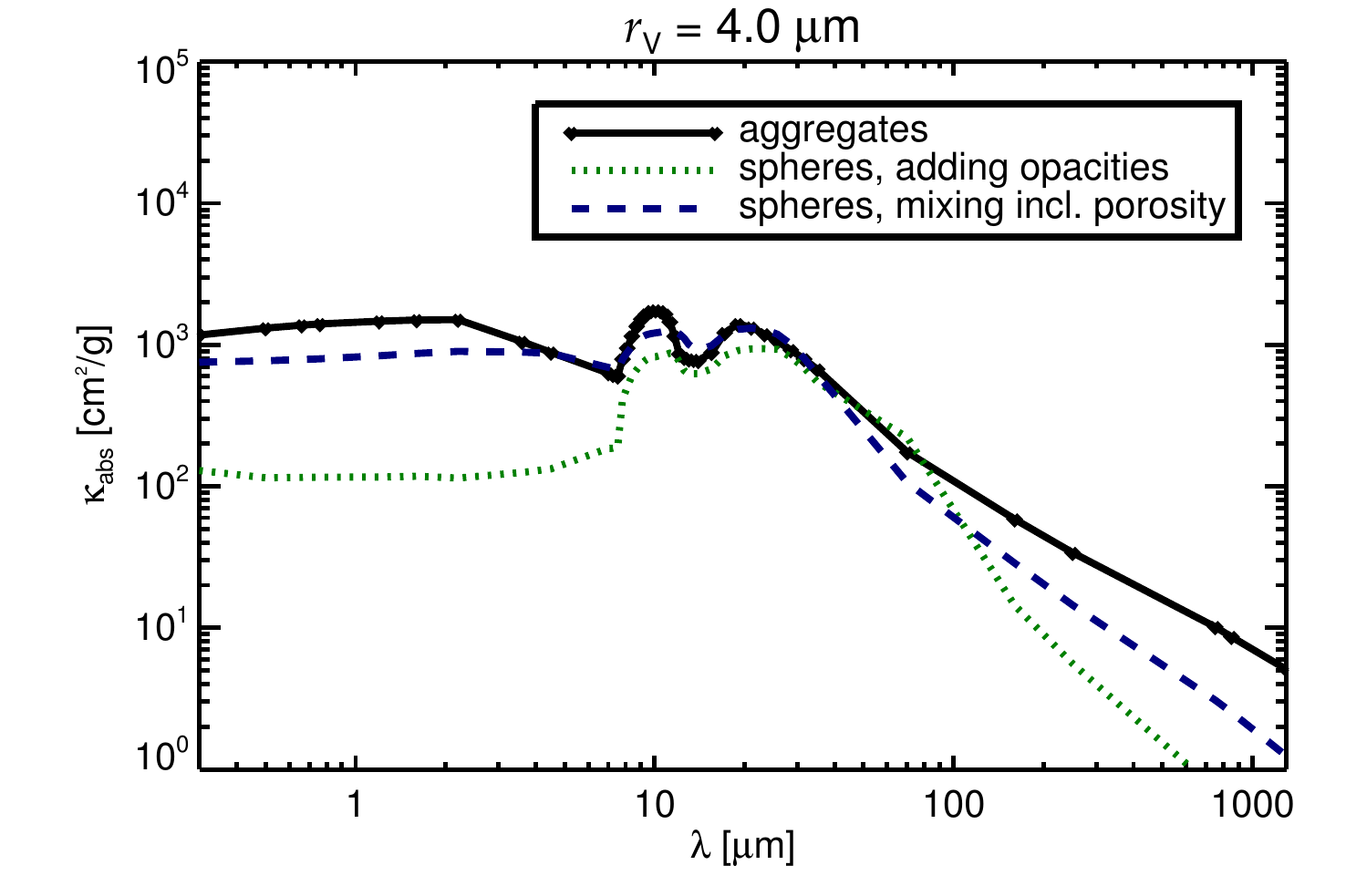}\includegraphics{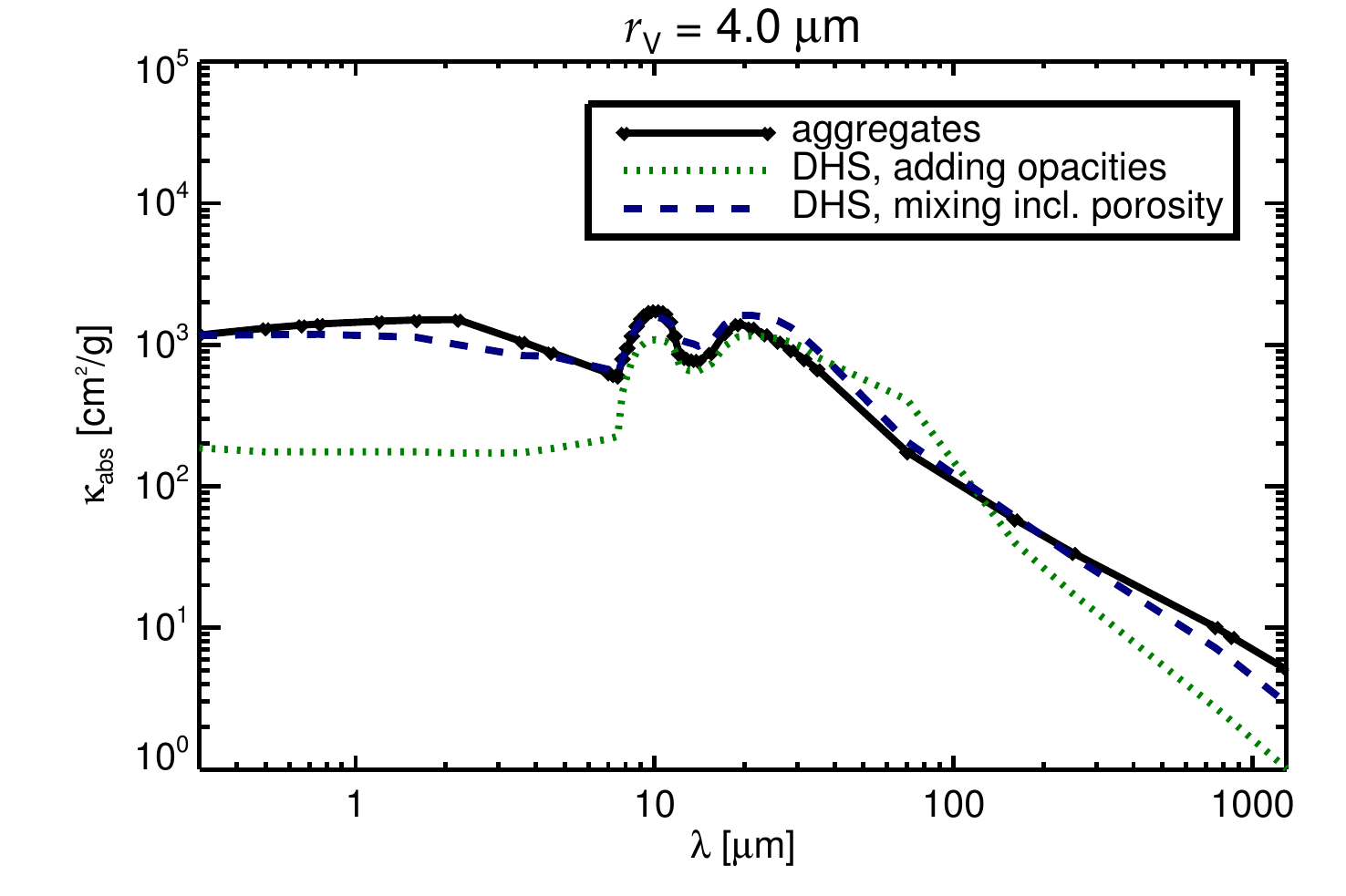}}}
\caption{Absorption cross section of the aggregate particles as a function of wavelength for three different particle sizes compared to the approximate methods. The upper panels show the aggregates for a volume-equivalent radius of $0.2\,\mu$m, the
middle panels show$1.2\,\mu$m, and the lower panels $4.0\,\mu$m particles. The left panels show spherical particles with and without added porosity, the right panels show the opacities computed with the DHS method with and without porosity.}
\label{fig:absorption spectra}
\end{figure*}

\subsection{Cross sections}
\label{sec:crosssections}

The absorption, scattering, and extinction cross sections of the aggregate particles are plotted in Fig.~\ref{fig:crosssections} for a few selected aggregate sizes. In general, the opacity decreases when the aggregate size increases. This occurs first at the shortest wavelengths. In addition, it is clear that scattering becomes increasingly important when the aggregate size increases. For aggregate sizes of a few micron, scattering is already a major component even at mid-infrared wavelengths -- an important effect for multiwavelength radiative transfer modeling of protoplanetary disks where aggregates of these sizes are expected to be very common.

When we compare the cross sections for absorption computed with DDA to those computed using the approximate methods (Fig.~\ref{fig:absorption spectra}), we see a number of important effects.

First of all, it is clear that adding the opacities of single, homogeneous particles is not a good approximation in the optical part of the spectrum for aggregates. In the upper panels, which are basically the unaggregated constituent monomers, this approximation seems to work reasonably well, at least at wavelengths lower than 20$\mu$m.  However, when aggregation starts (the two lower rows in Fig.~\ref{fig:absorption spectra}), this approximation clearly underestimates the absorption cross section. We speculate that the reason for this is that in this approximation, each homogeneous particle itself is assumed to be of the size of the simulated aggregate. For the lower panels we therefore have a $4\,\mu$m silicate grain, a $4\,\mu$m carbonaceous grain, and a $4\,\mu$m iron sulfide grain. In the real aggregate, however, the separate material components each contribute a small fraction so they cannot combine to such large particles. This causes the opacity of the aggregate particles to be much higher. When using
the effective medium theory, this effect is approximately simulated and the opacities match much better. Additionally, the porosity included in these computations causes the grains to have a higher opacity.

\begin{figure*}[!t]
\centerline{\resizebox{0.95\hsize}{!}{\includegraphics{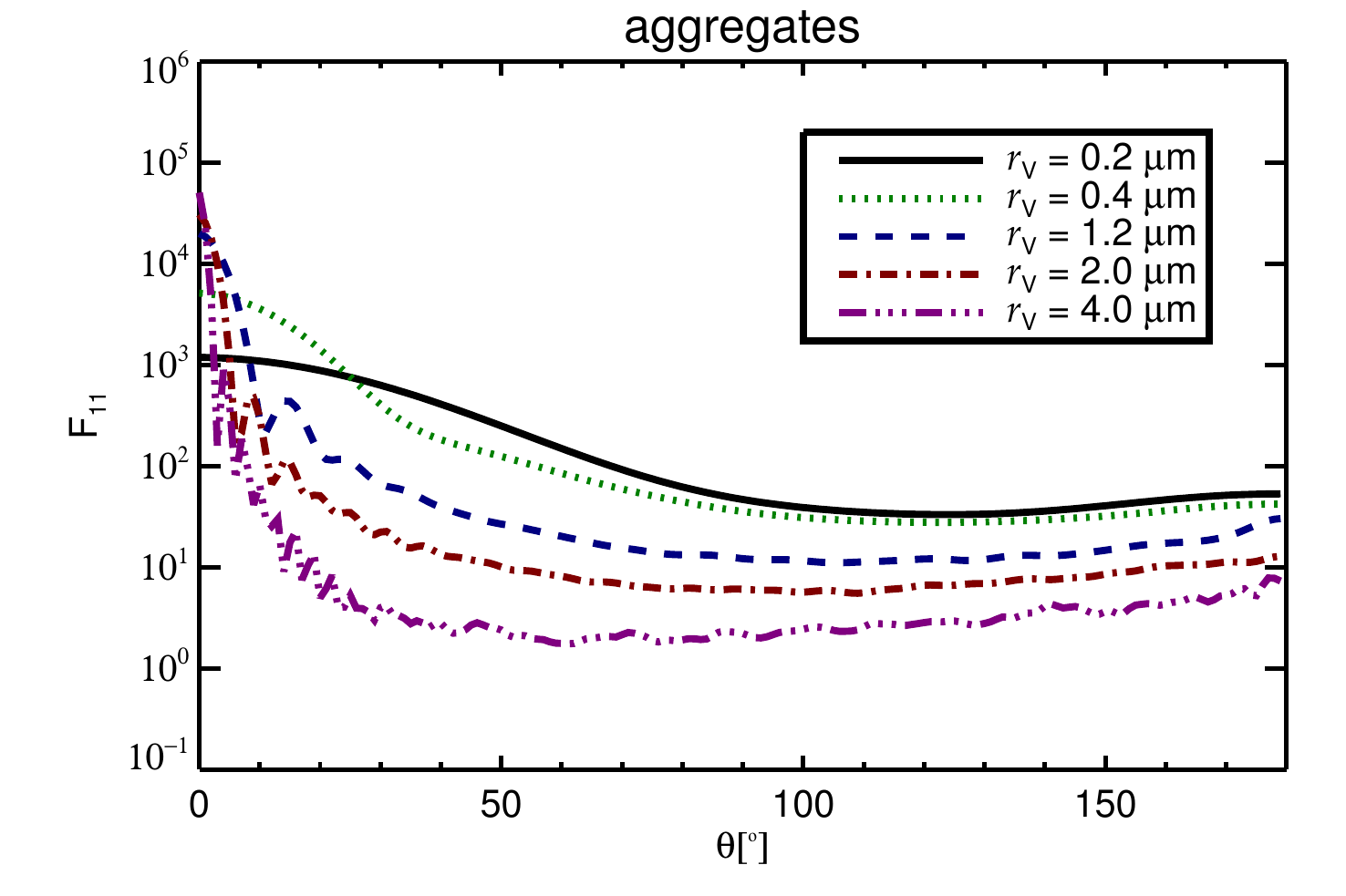}\includegraphics{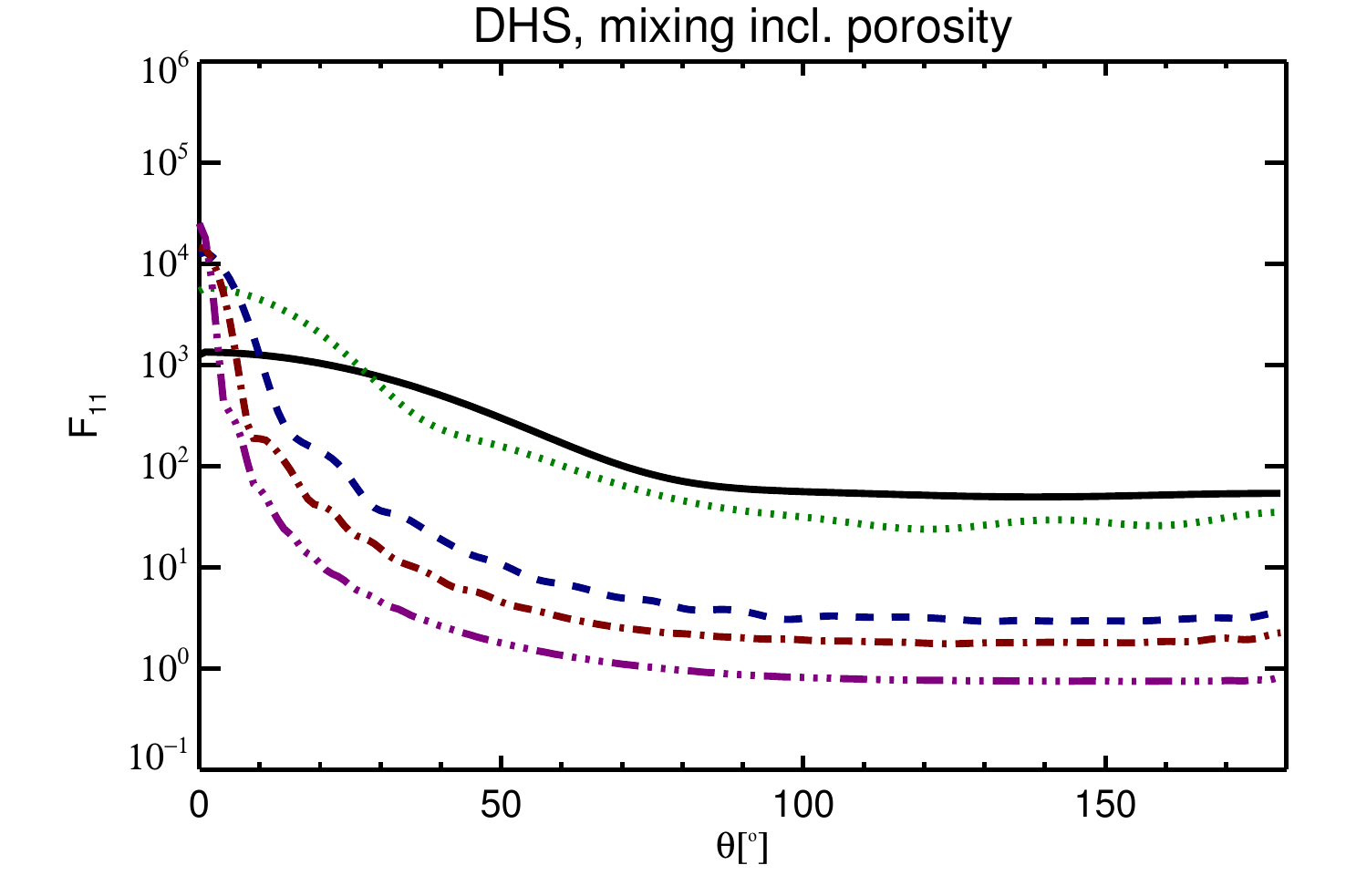}}}
\caption{Phase function for scattering at $\lambda=0.55\,\mu$m. The left panel shows the aggregates of different sizes. The right panel shows the DHS model with 25\% porosity. The other approximate methods show very similar phase functions.}
\label{fig:F11}
\end{figure*}

The second important result is the opacity at mm wavelengths. The slope is much better simulated using the effective medium mixing models than by adding opacities. The absolute value is much better represented using the DHS model compared to the homogeneous spheres. The best approximation for the millimeter opacity, both slope and absolute value, is when using the DHS model in combination with effective medium theory.  We note that even this method still has a significantly steeper slope than the full aggregate computations.

The opacity at 1300$\mu$m is for the medium-size aggregates $\kappa_\mathrm{abs}=5.1\,$cm$^2$/g. This compares reasonably well to the value used for example by \citet{2005ApJ...631.1134A} of $\kappa_\mathrm{abs}=3.5\,$cm$^2$/g. The opacity at millimeter wavelengths is fairly well approximated by a power law, $\kappa_\mathrm{abs}\propto\lambda^{-\beta}$. We find that for the aggregate computations $\beta\sim1.2$. The DHS with effective medium mixing has a slope of $\beta\sim1.6$.
However, these values are very sensitive to the abundance of carbon and iron sulfide and to the exact composition of the carbon. Preliminary computations we did for more conducting forms of carbon yielded an opacity that can be orders of magnitude higher and a slope that is much more shallow. This can have very important consequences for the interpretation of the millimeter slopes of protoplanetary disk spectra, since a shallow slope is usually interpreted as a sign of grain growth. We consider the discussion of the exact value of the millimeter opacity and its slope beyond the scope of this paper, but do note here that the DHS model with effective medium theory reproduces the millimeter opacity of the DDA aggregate computations, in contrast to the other approximations we tested. This suggests that DHS is a good approximation to use when studying the effects of composition on the millimeter opacity and slope.

It is also clear from the curves in Fig.~\ref{fig:absorption spectra}
that the 10$\,\mu$m silicate feature, which is frequently used as an indicator of
particle size, is not well represented by the methods without
porosity. We also computed opacities using the effective medium theory,
but without the 25\% porosity.  This experiment yielded too weak
silicate features for all particle sizes larger than ~$1\,\mu$m.  This
solid-state feature seems to correlate to the overall porosity of the
particle. For compact particles the size dependence of the feature is
too strong. It is expected that for particles more fluffy than the one
we used, this effect is much stronger still, as has been shown by
\citet{2006A&A...445.1005M} for aggregates of different fractal
dimension.

\subsection{Scattering properties}

\begin{figure*}[!t]
\centerline{\resizebox{\hsize}{!}{\includegraphics{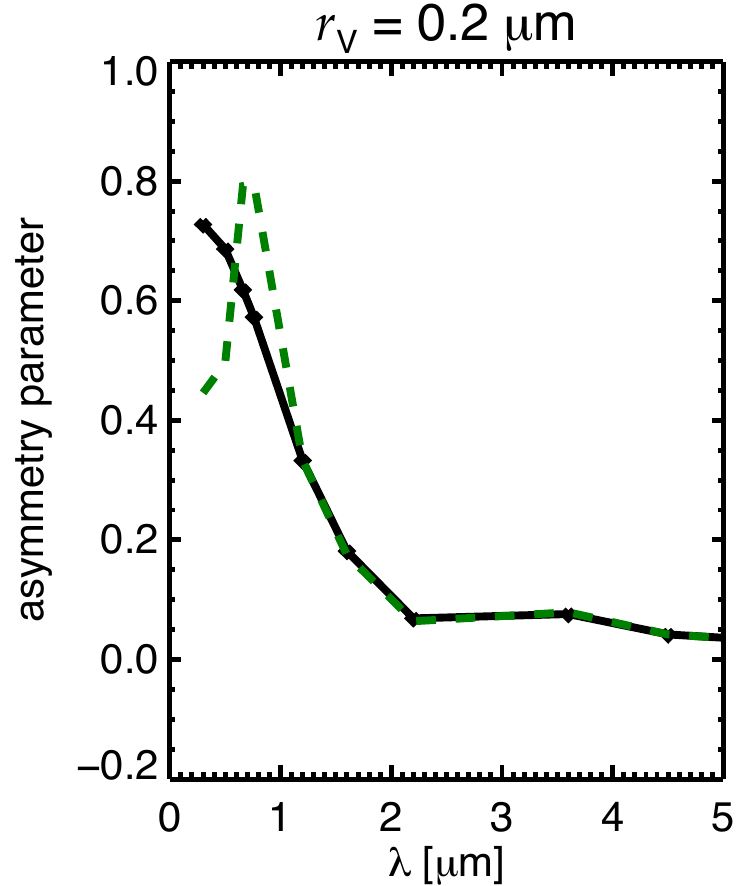}\includegraphics{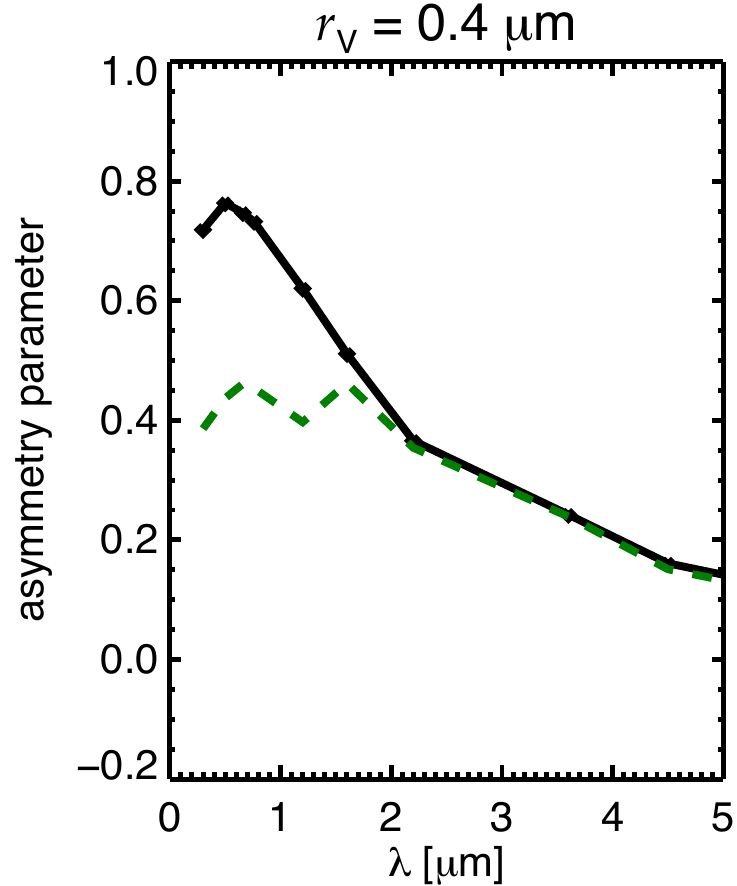}\includegraphics{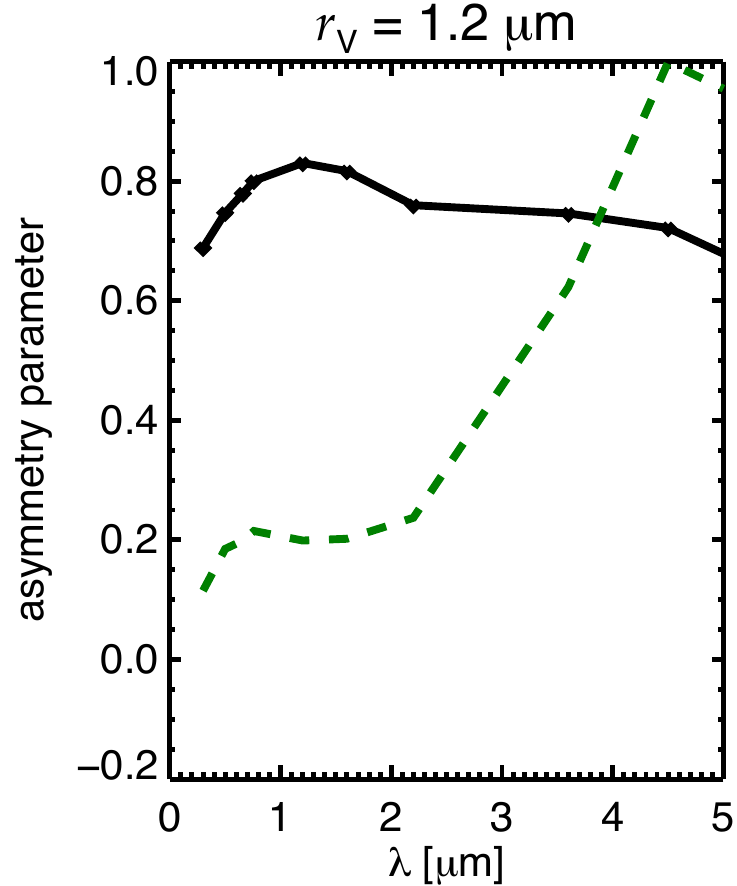}\includegraphics{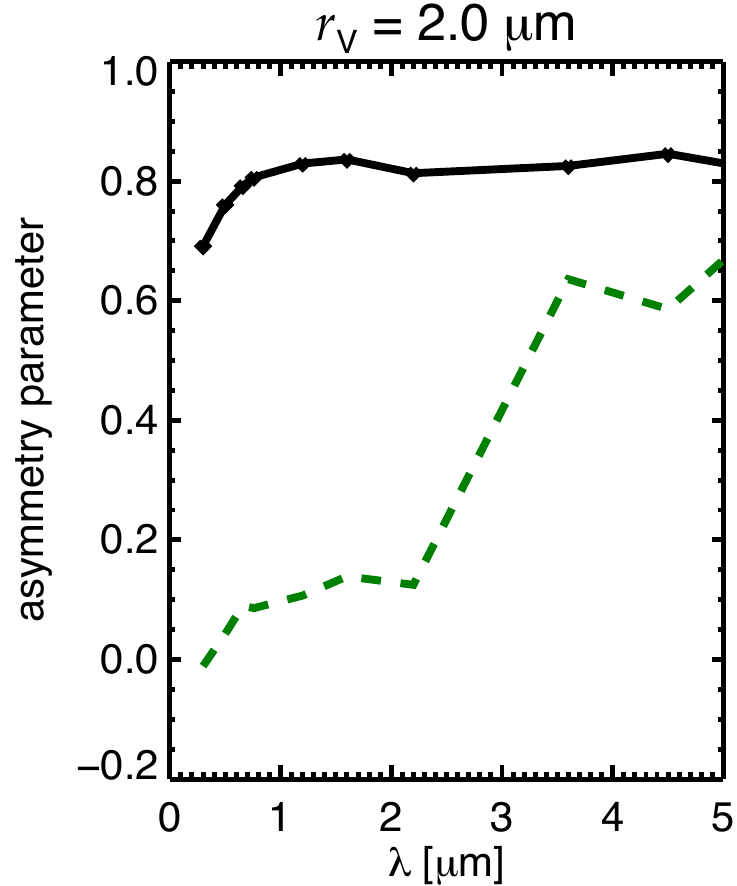}\includegraphics{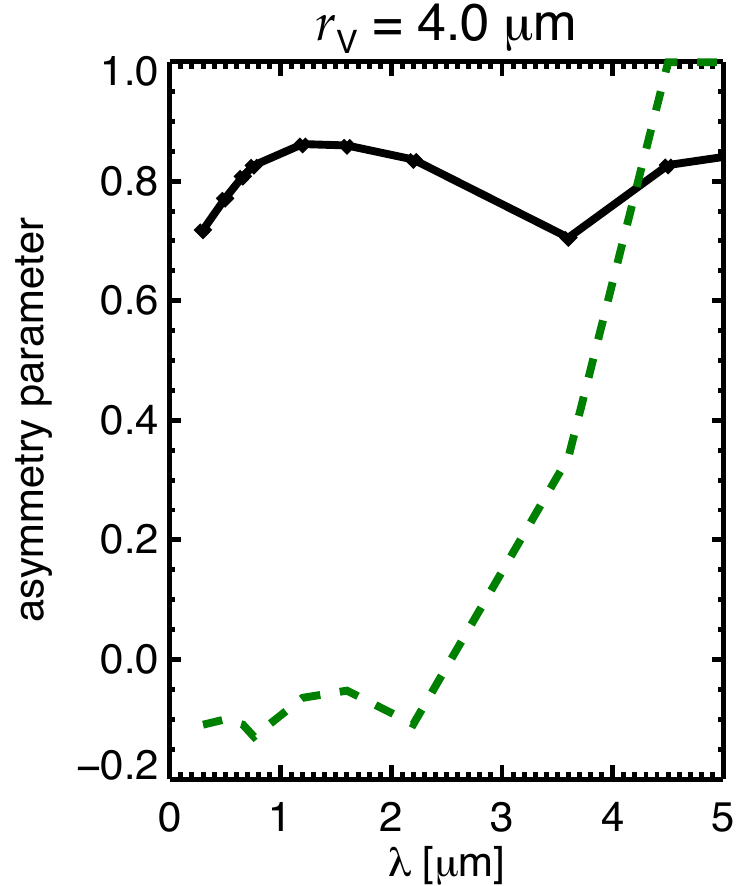}}}
\vspace{0.3cm}
\centerline{\resizebox{\hsize}{!}{\includegraphics{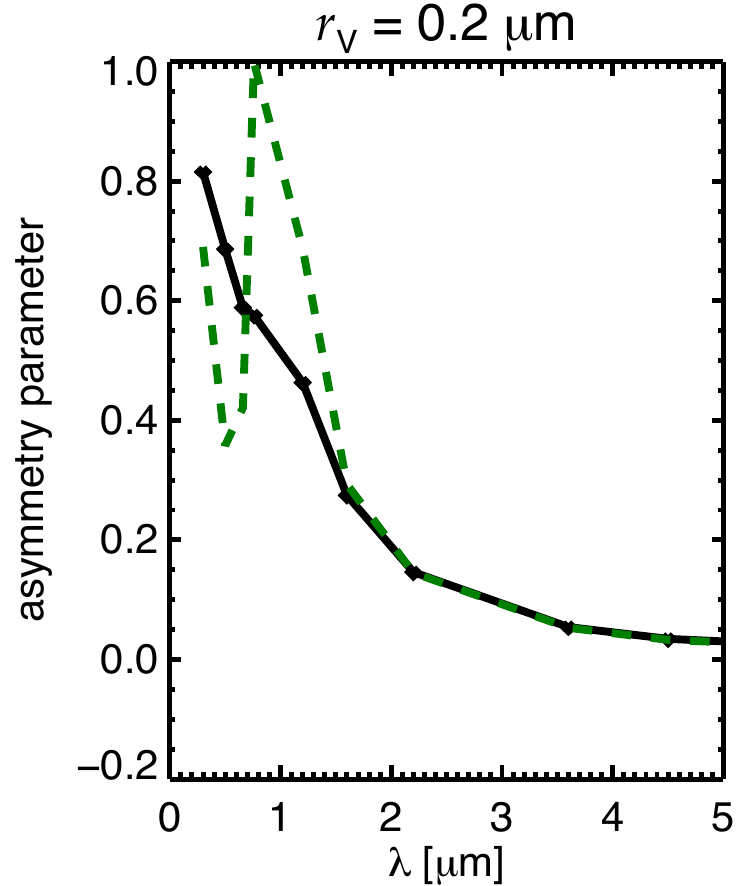}\includegraphics{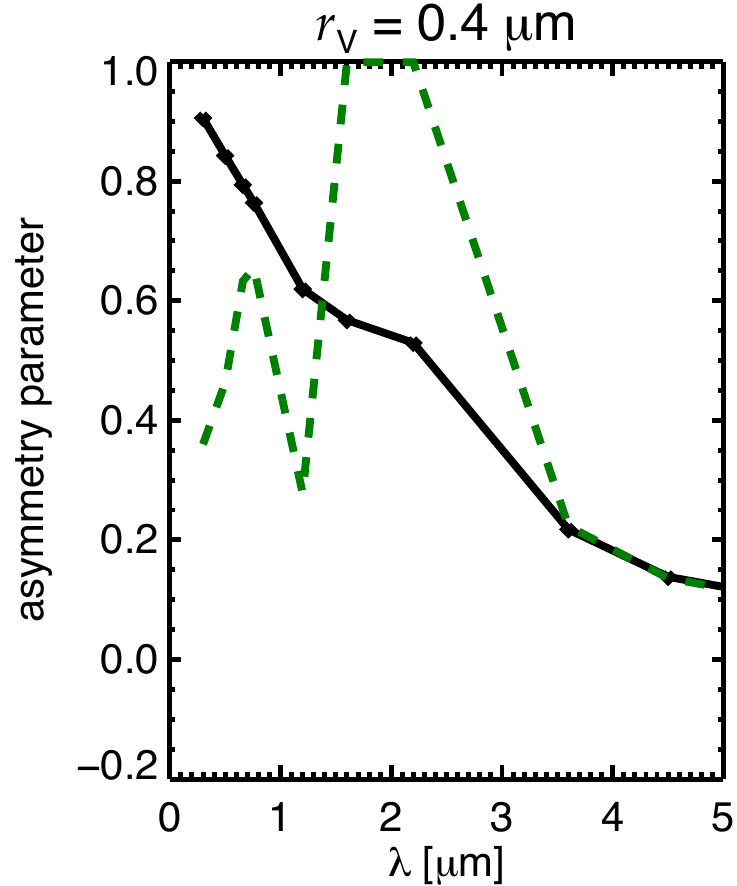}\includegraphics{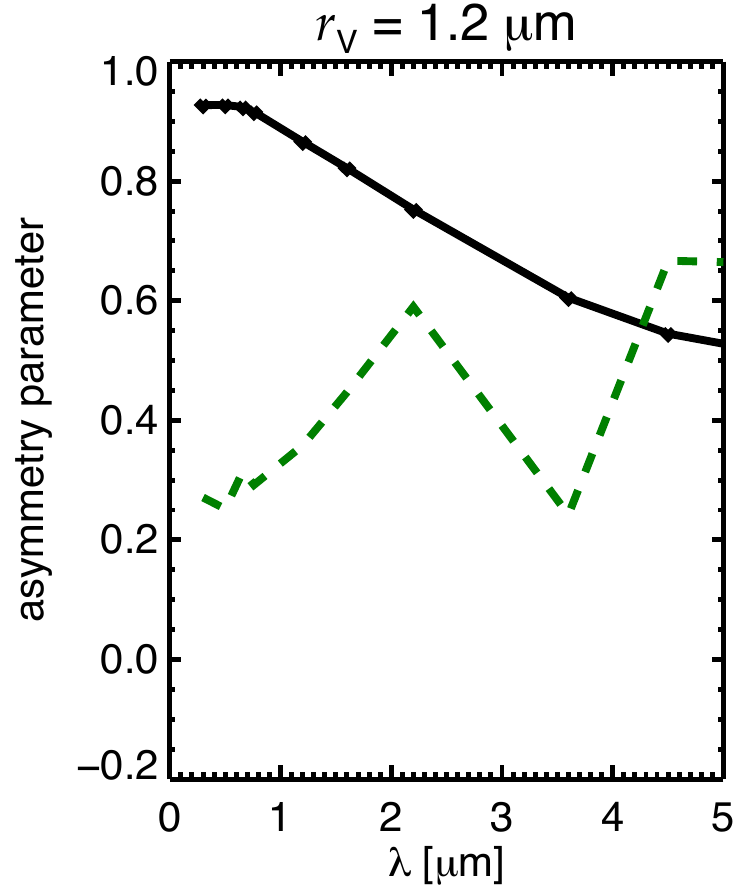}\includegraphics{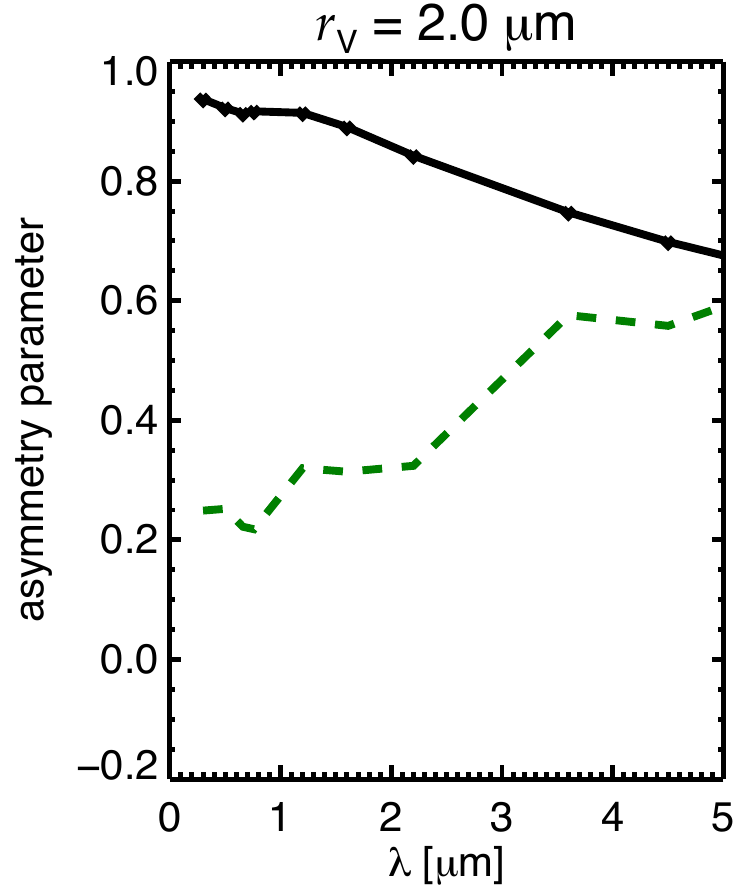}\includegraphics{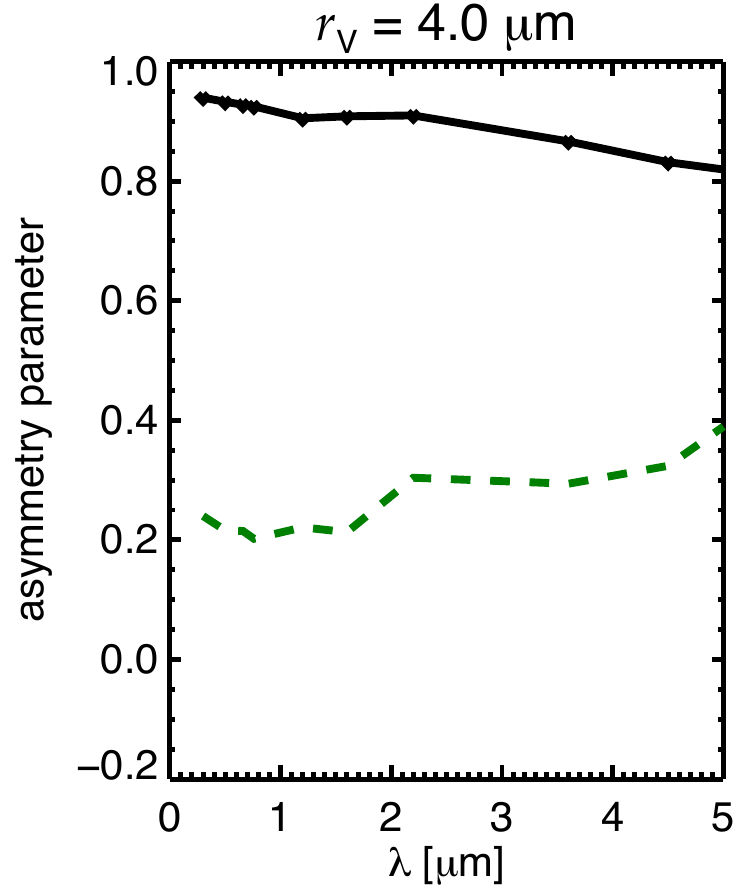}}}
\caption{Wavelength-dependent asymmetry parameter of the aggregate particles (upper panels) and the DHS distribution with porosity (lower panels) for different particle sizes. We show the formal asymmetry parameter of the entire phase function with solid black
lines and the effective asymmetry parameter with dashed green
lines. This was obtained by fitting a Henyey-Greenstein phase function to the central $80^\circ$ of the phase function, i.e., from $50 - 130^\circ$ scattering angle, representative of a disk with inclination of $40^\circ$.}
\label{fig:asymmetry}
\end{figure*}

\subsubsection{Phase function}

In Fig.~\ref{fig:F11} we show the phase functions, $F_{11}$, computed for scattering at a wavelength of $\lambda=0.55\,\mu$m. The overall shape of this curve is clearly similar to the simplified approach. Especially the forward-scattering part of the phase function, up to $\sim50^\circ$ scattering, is very similar. From this scattering angle toward more backscattering directions, the phase functions of the real aggregates do differ from all simplified methods. The aggregates  show a much flatter and even slightly \textup{\textup{\textup{\emph{\textup{increasing}}}} }phase function with scattering angle, where all simplified methods display a \emph{\textup{decreasing}} phase function with scattering angle. The reason for this is the concaveness of the aggregates outer layers. The approximate methods are all very smooth particles with a convex and completely smooth surface. For large particles, this causes strong forward Fresnel reflections. However, for a concave or rough surface, a behavior much more in line with that also outlined in \citet{2010A&A...509L...6M} is expected. These relatively small aggregates already show the start of the backward-scattering behavior expected from simple geometrical optics (like the lunar phases). Indeed, the shape of the phase function of the $4\,\mu$m aggregates can be completely understood from a superposition of a diffraction part and a backward-reflecting curve computed from regolith reflectance theory \citep{1981JGR....86.3039H}. This effect is expected to be especially strong for more compact particles. By increasing the aggregate fluffiness, the effect will eventually disappear.

In Fig.~\ref{fig:asymmetry} we plot the asymmetry parameter of the particles as a function of wavelength. The asymmetry parameter, $g=\left<\cos\theta\right>$ , is defined as\begin{equation}
\label{eq:asymmetry}
g=\frac{\int_{4\pi} F_{11}(\theta) \cos\theta~d\Omega}{\int_{4\pi} F_{11}(\theta)~d\Omega}.
\end{equation}

In observations, we often cannot access the entire phase function, and conclusions are drawn about the asymmetry parameter from a limited range of scattering angles. 
For example, debris disks are optically thin and $F_{11}$ can be measured directly from scattered-light images. If the disk is resolved, then we know the disk inclination and the range of scattering angle probed. The scattering properties then have to be compared over the same range of scattering angles.
To examine the true diagnostic value of this range, we also show in Fig.~\ref{fig:asymmetry} the asymmetry parameter that would
be derived when only the central $80^\circ$ of scattering can be accessed, for example, for a debris disk with inclination $40^\circ$ from face-on. As is often done in cases like this, we fit a Henyey Greenstein phase function,
\begin{equation}
\label{eq:HG}
F_{11}^\mathrm{HG}\propto\frac{1-g^2}{(1-2g\cos\theta+g^2)^{3/2}},
\end{equation}
to this part of the computed phase function.  This yields an observed, or effective, asymmetry parameter of that part of the phase function.  It is plotted in Fig.~\ref{fig:asymmetry} .  The effective asymmetry parameter can be very different from the formal one computed by Eq.~\ref{eq:asymmetry}. For short wavelengths and large aggregates it can even be negative, corresponding to the part of the phase function that increases with scattering angle, that is, the dominant backward-scattering part.  In real observations, the orientation of the system is often unknown
and only the absolute value of the asymmetry parameter can be obtained. Only whith additional information on the near and far side of the disk can the sign be obtained. It is expected that the wavelength region with dominant backward scattering (effective asymmetry parameter $g<0$) becomes broader with increasing aggregate size. The range of scattering angles over which the phase function is backward scattering is also expected to grow with increasing aggregate size because the diffraction peak, which is heavily forward scattering, moves to smaller scattering angles. This effect is not captured by any of the approximate methods.

\subsubsection{Effective albedo}

As discussed in the previous section, we often cannot access the full range of scattering angles in an observed object. In the case of anisotropic scattering, this can also cause a large part of the scattered light to be scattered away from the observer. Especially in the case of extreme forward scattering, this can cause the observed scattered-light flux to be much lower than what would be expected by considering the single-scattering albedo alone.  This effect was introduced by \citet{2013A&A...549A.112M} to explain the faint scattered-light halo of the Herbig star HD~100546 and its red color.

\begin{figure*}[!t]
\centerline{\resizebox{\hsize}{!}{\includegraphics{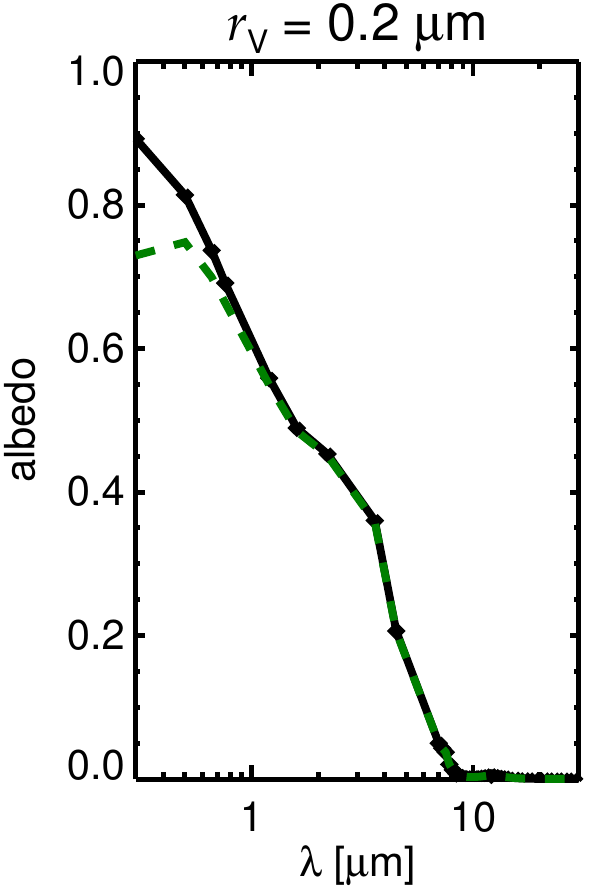}\includegraphics{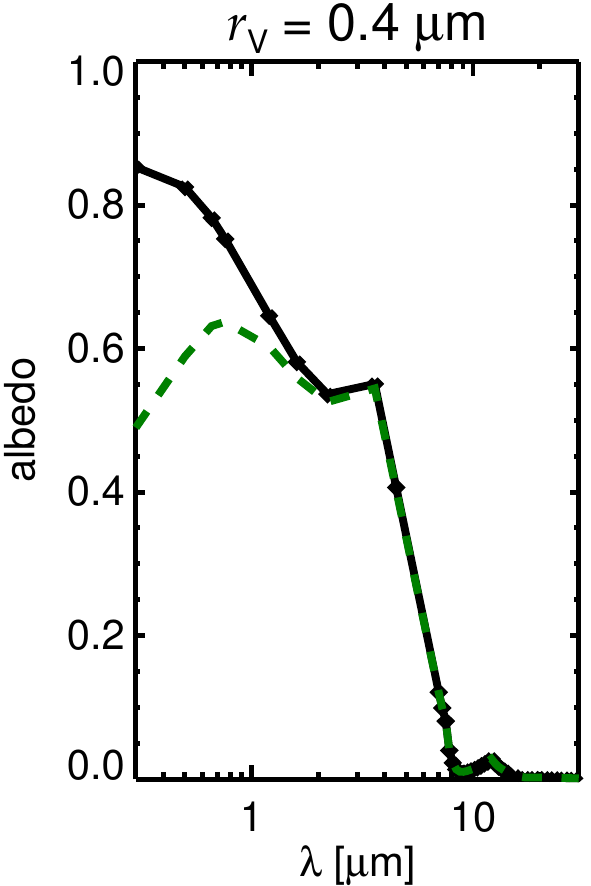}\includegraphics{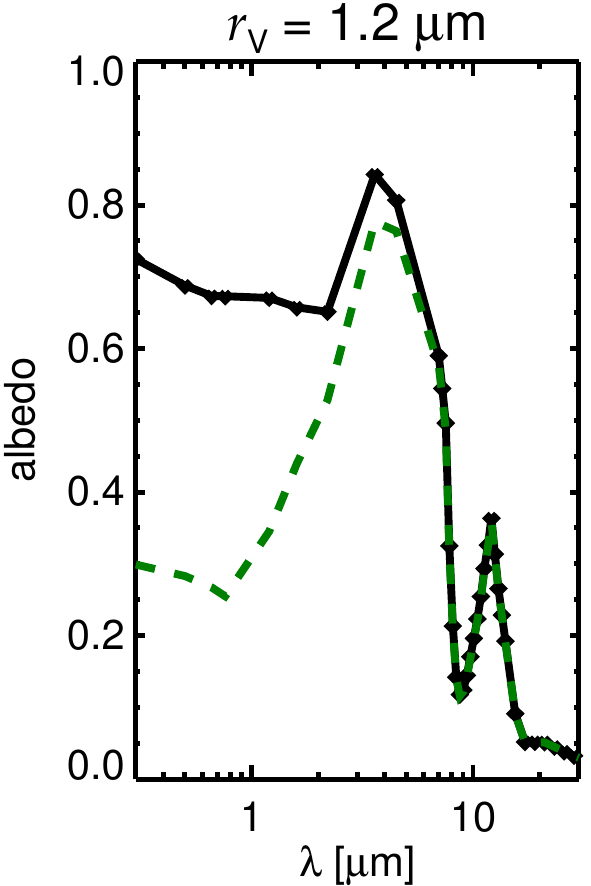}\includegraphics{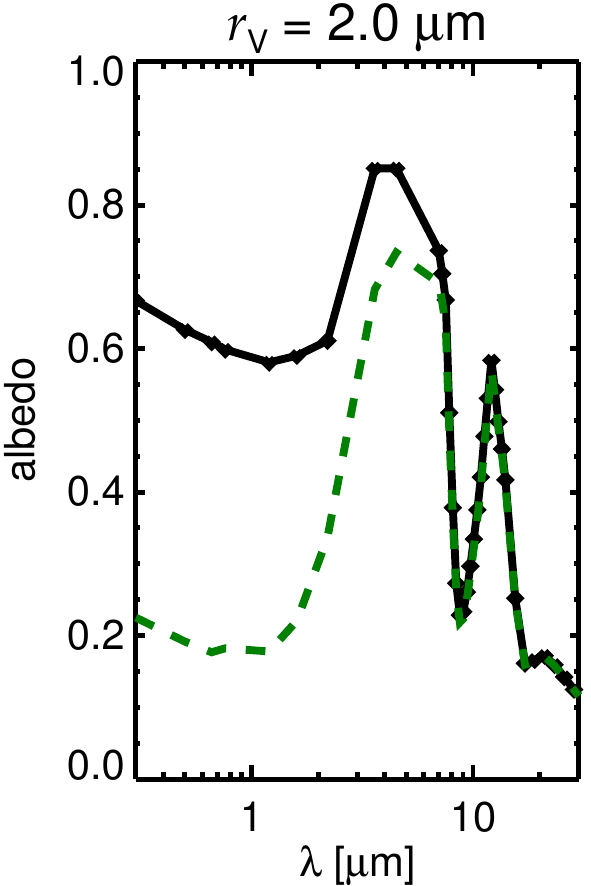}\includegraphics{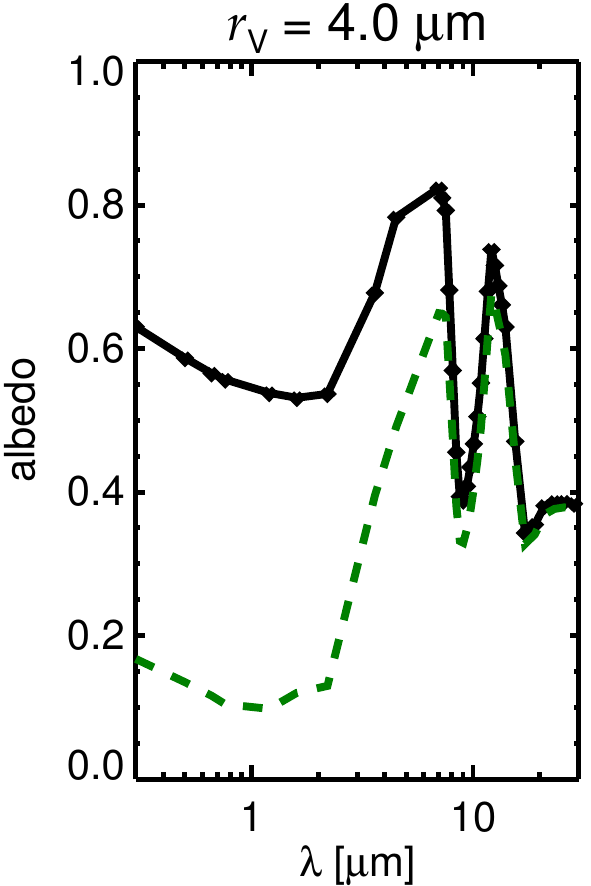}}}
\vspace{0.3cm}
\centerline{\resizebox{\hsize}{!}{\includegraphics{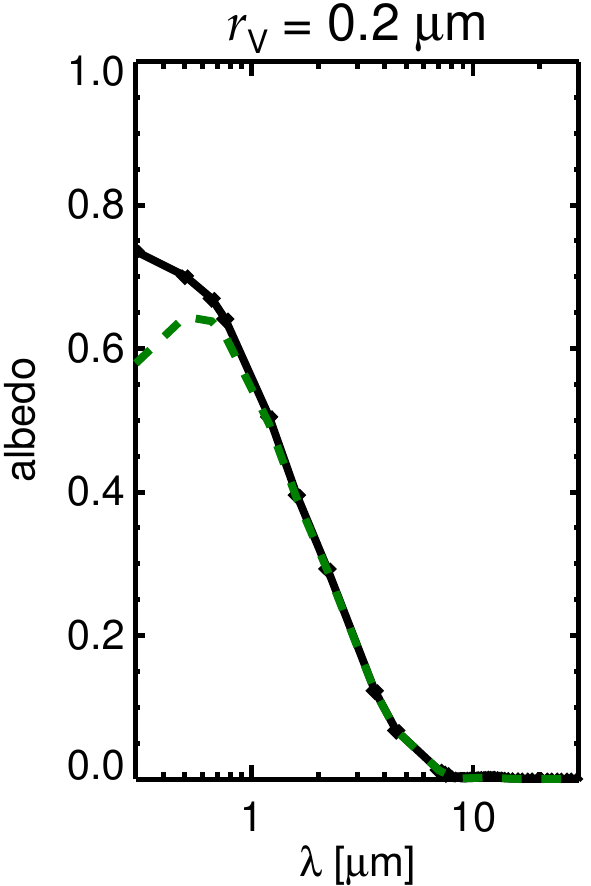}\includegraphics{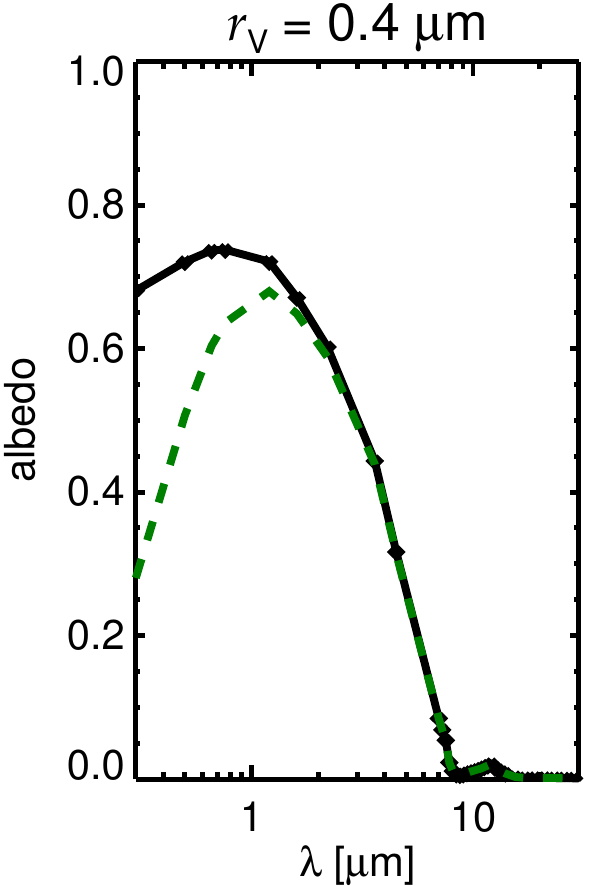}\includegraphics{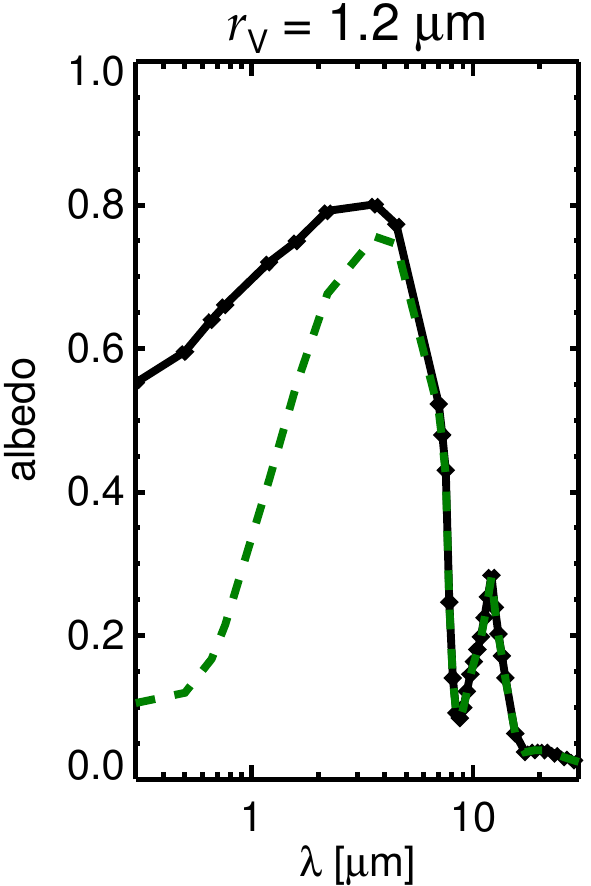}\includegraphics{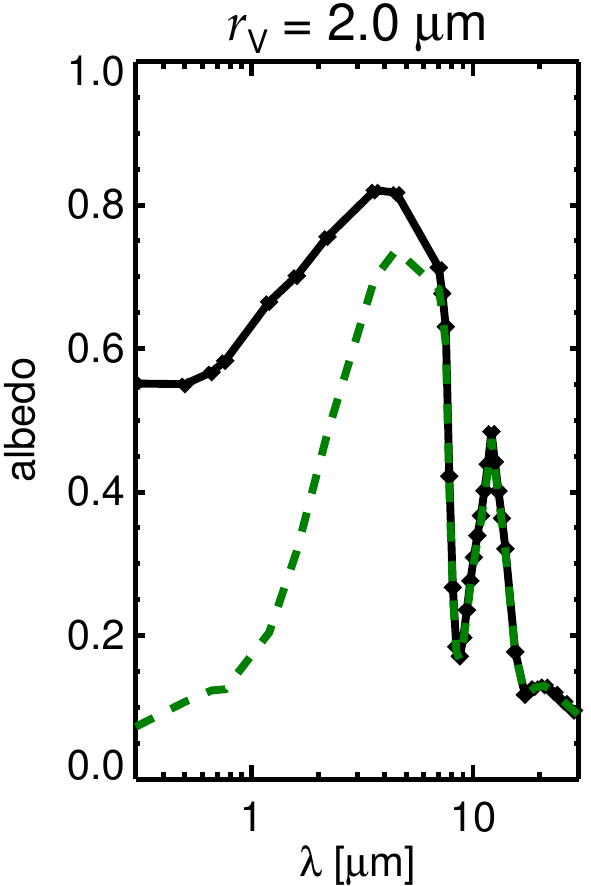}\includegraphics{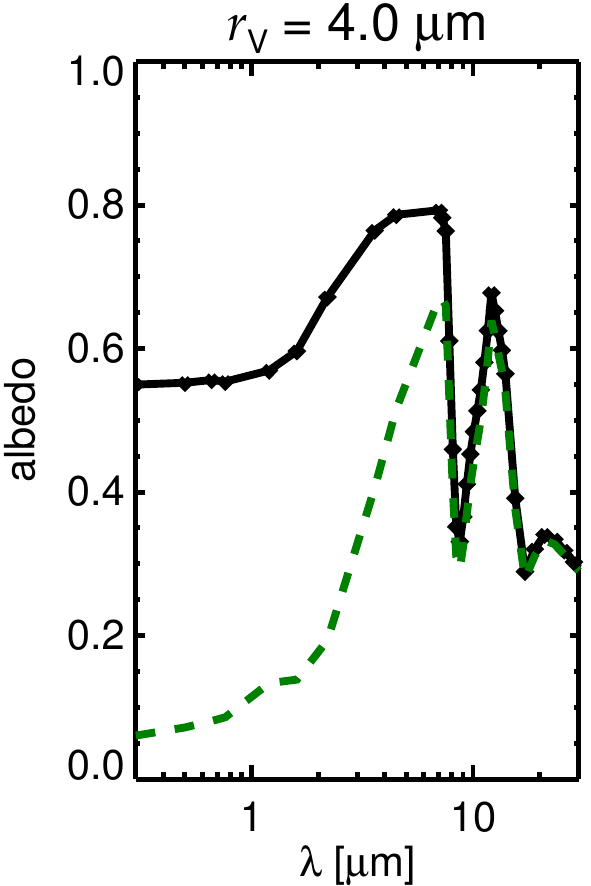}}}
\caption{Wavelength-dependent (effective) albedo of the aggregate particles (upper panels) and the DHS distribution with porosity (lower panels) for different particle sizes. We show the total albedo with solid black lines and the effective albedo with dashed
green lines, i.e., removing the first 10$^\circ$ of scattering. The other approximate methods display a similar behavior, but with differently detailed shapes of the curves.}
\label{fig:albedo}
\end{figure*}

In Fig.~\ref{fig:albedo} we show the total and the effective albedo of the aggregates and the values corresponding to the simplified computations. The effective albedo is the albedo that would be computed if we were to cut out the forward $10^\circ$ of scattering. In observations the forward-scattering peak is often not seen, therefore the effective albedo provides a better measure for the observed albedo than the true albedo. The forward-scattering part of the phase function is usually very high, causing the effective albedo to be smaller than the true albedo in most cases.  The effect is strongest for shorter wavelengths, where the phase function is more asymmetric.  This wavelength dependence causes the effective scattering to be redder than what would be expected from isotropic scattering (the full albedo). The wavelength dependence and absolute value of the (effective) albedo are all captured equally well by the approximate methods.

\begin{figure*}[!tp]
\centerline{\resizebox{0.8\hsize}{!}{\includegraphics{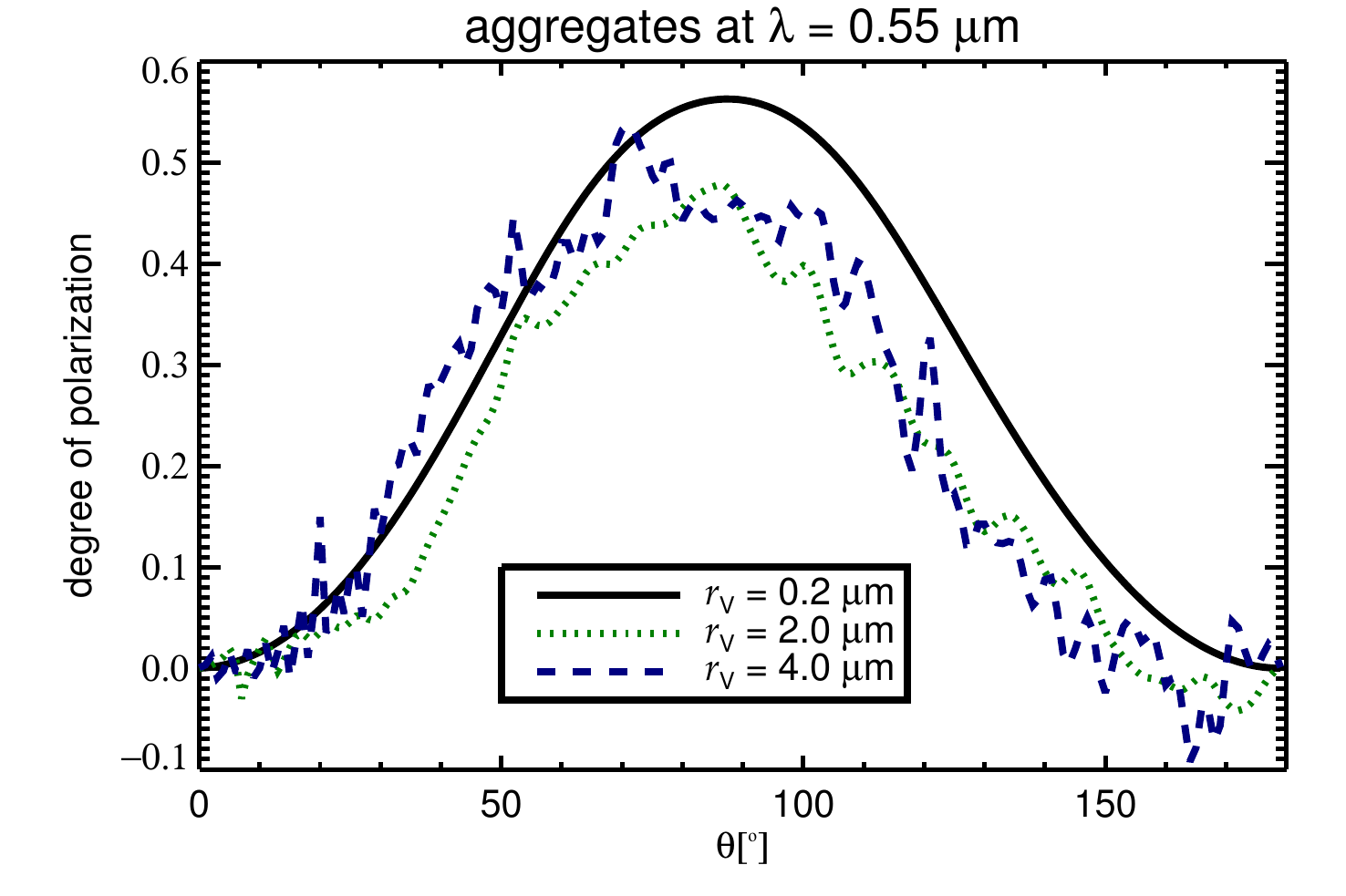}\includegraphics{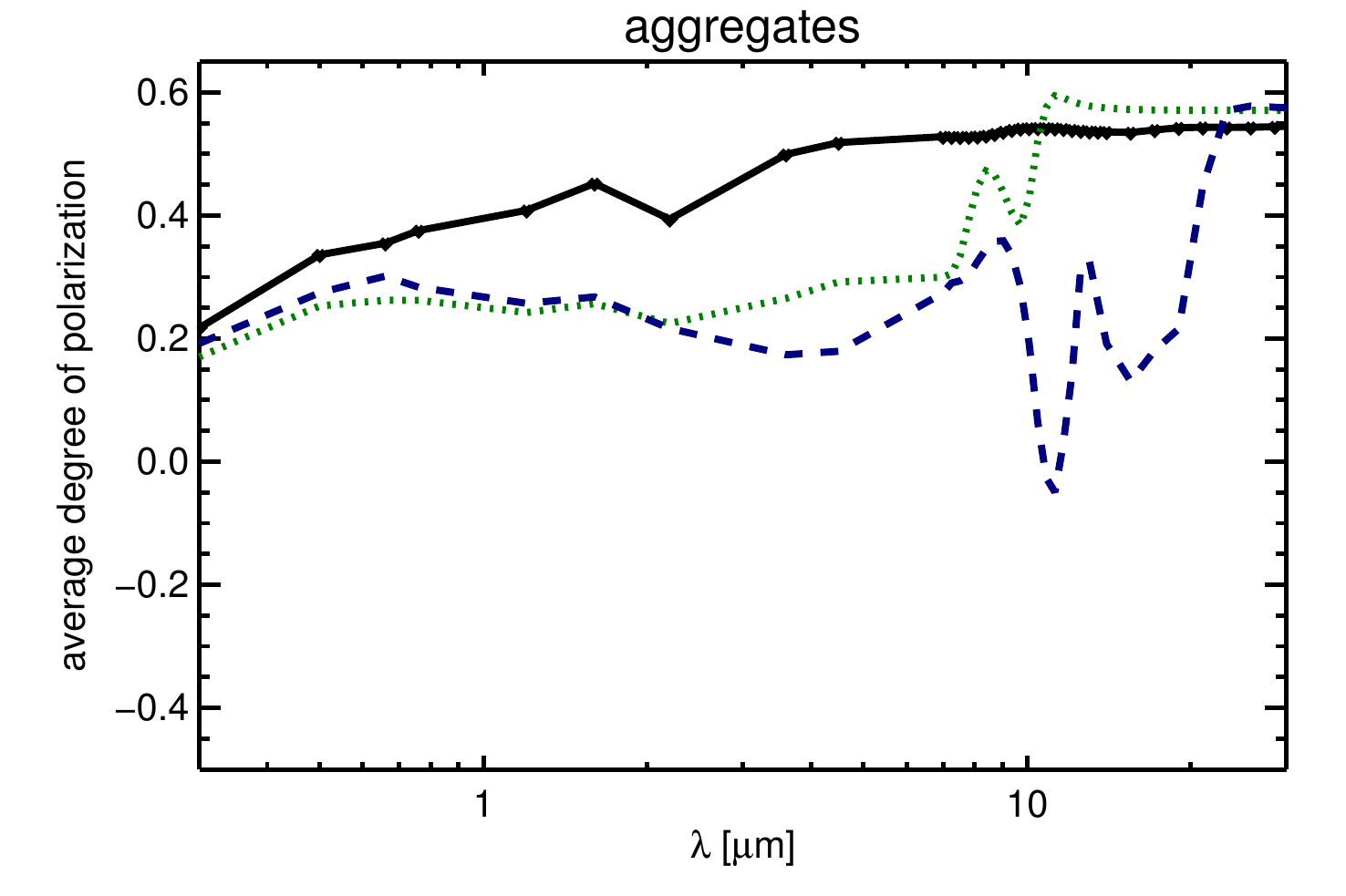}}}
\centerline{\resizebox{0.8\hsize}{!}{\includegraphics{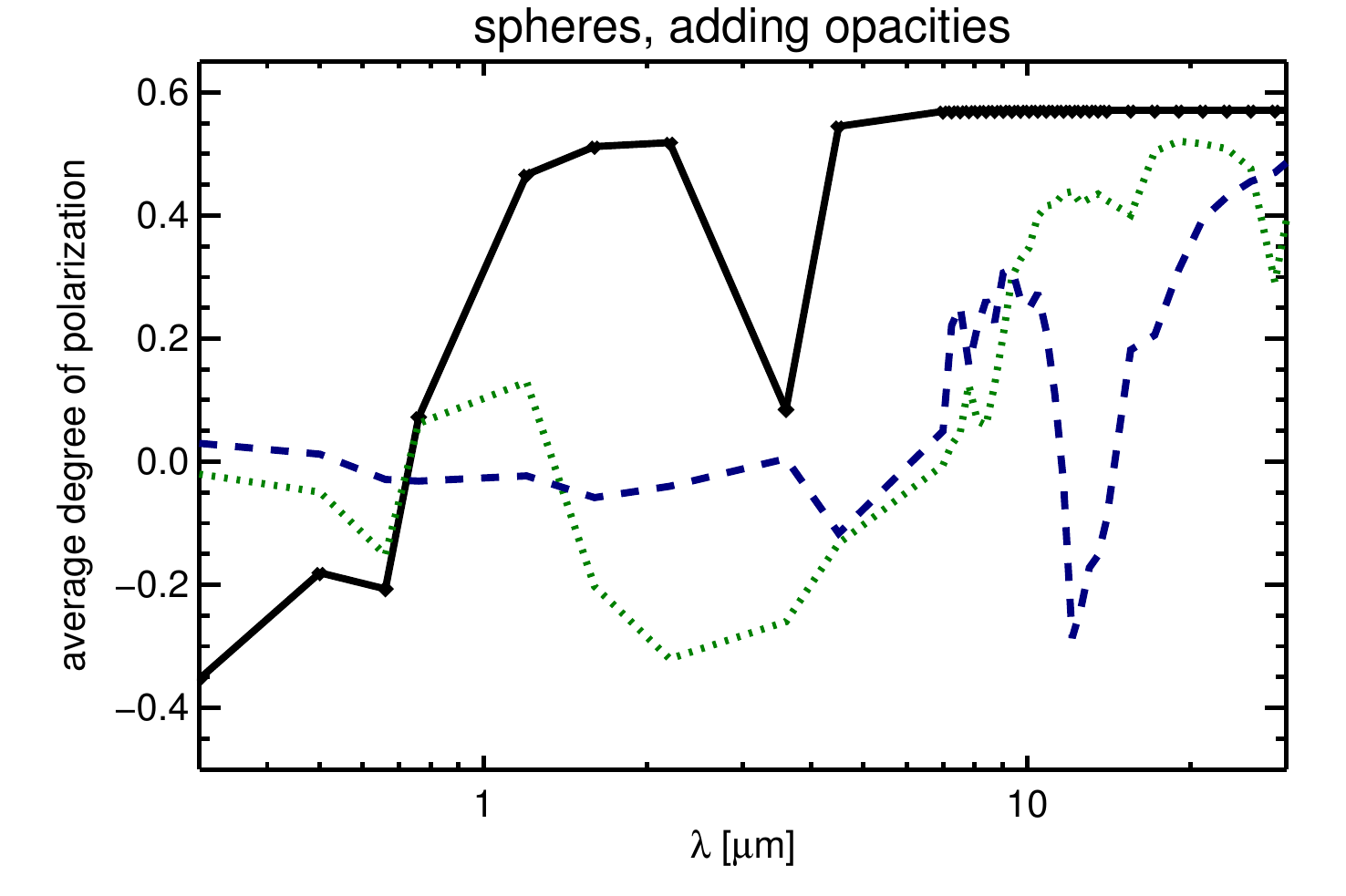}\includegraphics{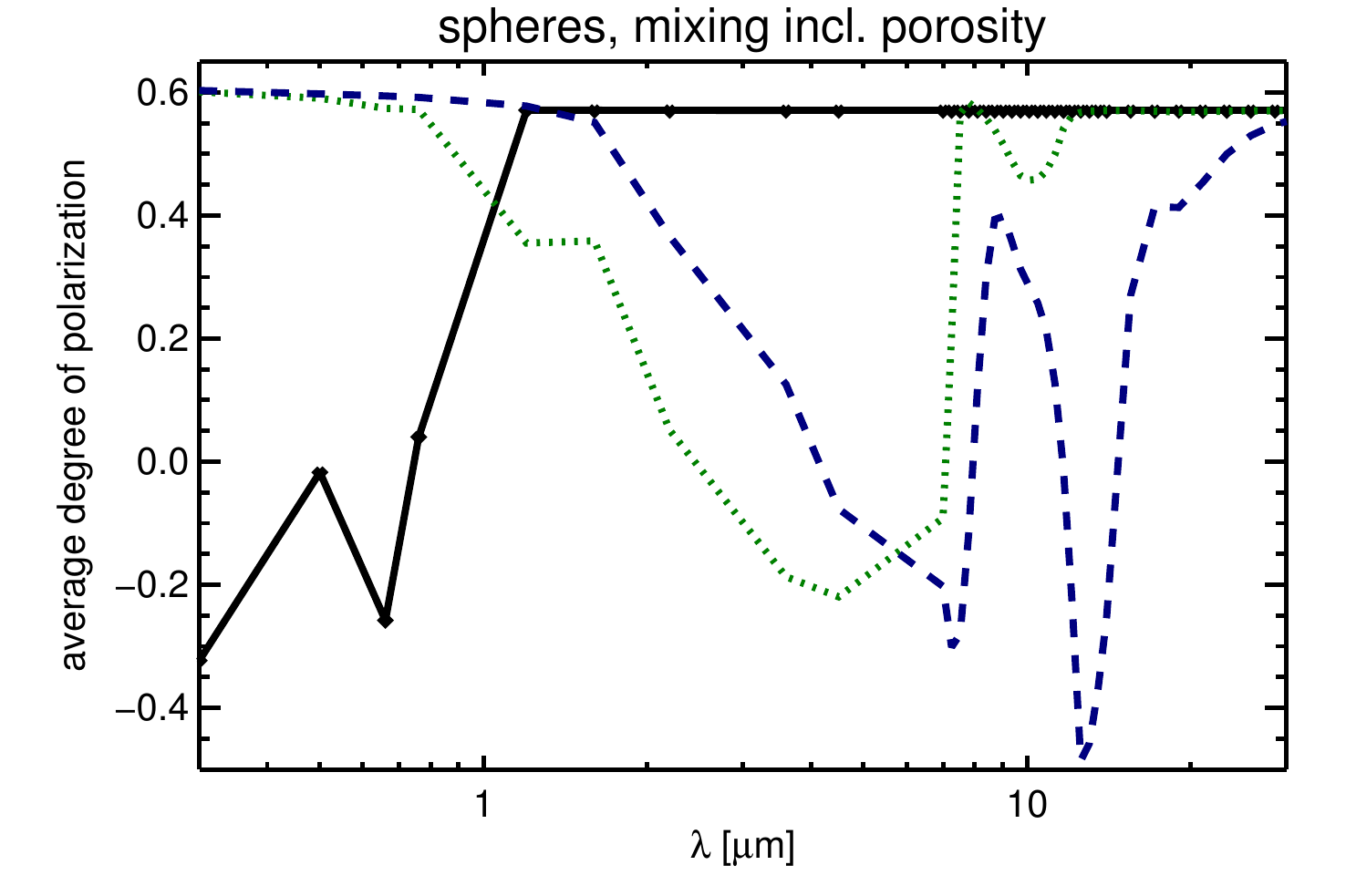}}}
\centerline{\resizebox{0.8\hsize}{!}{\includegraphics{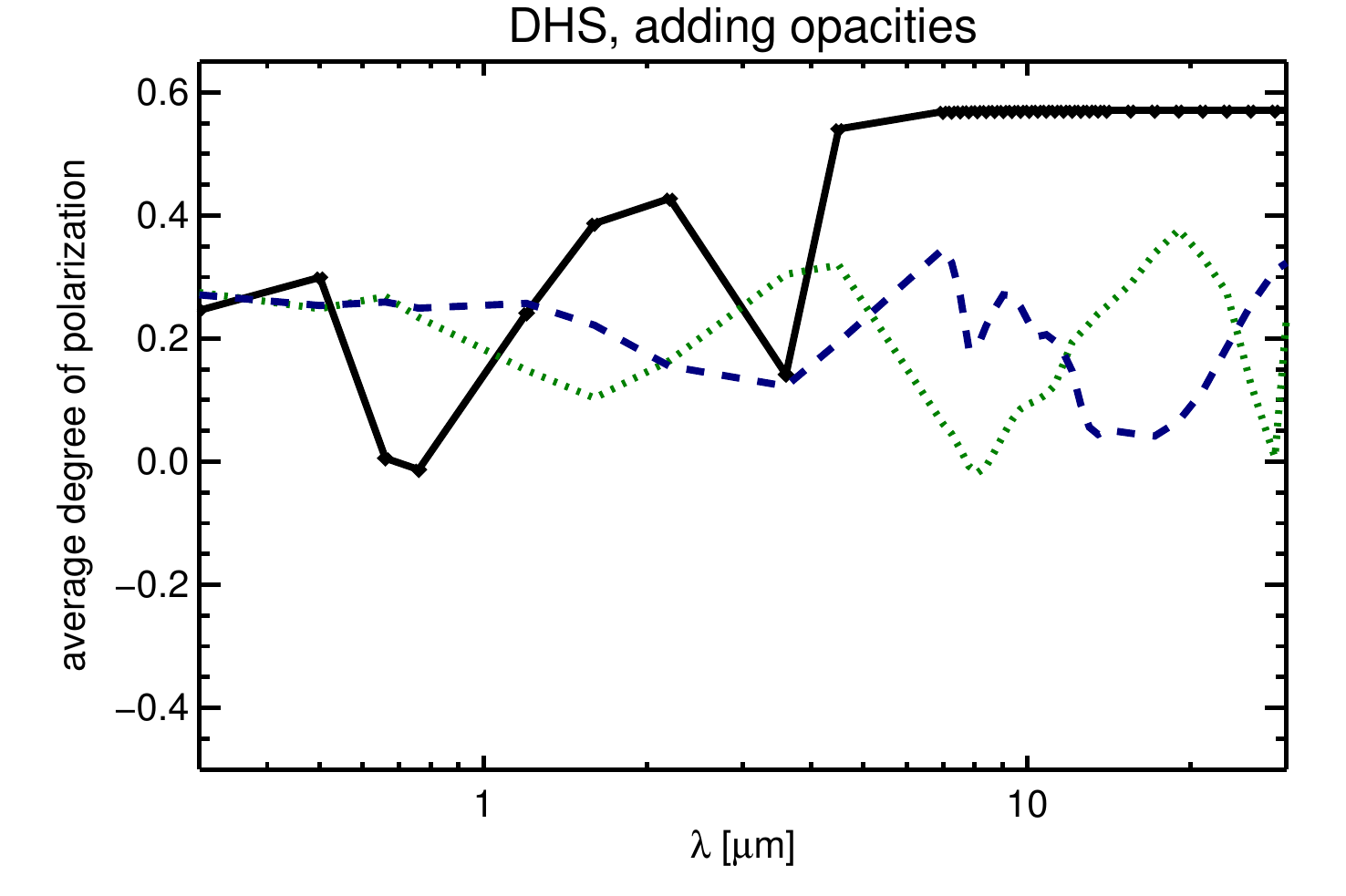}\includegraphics{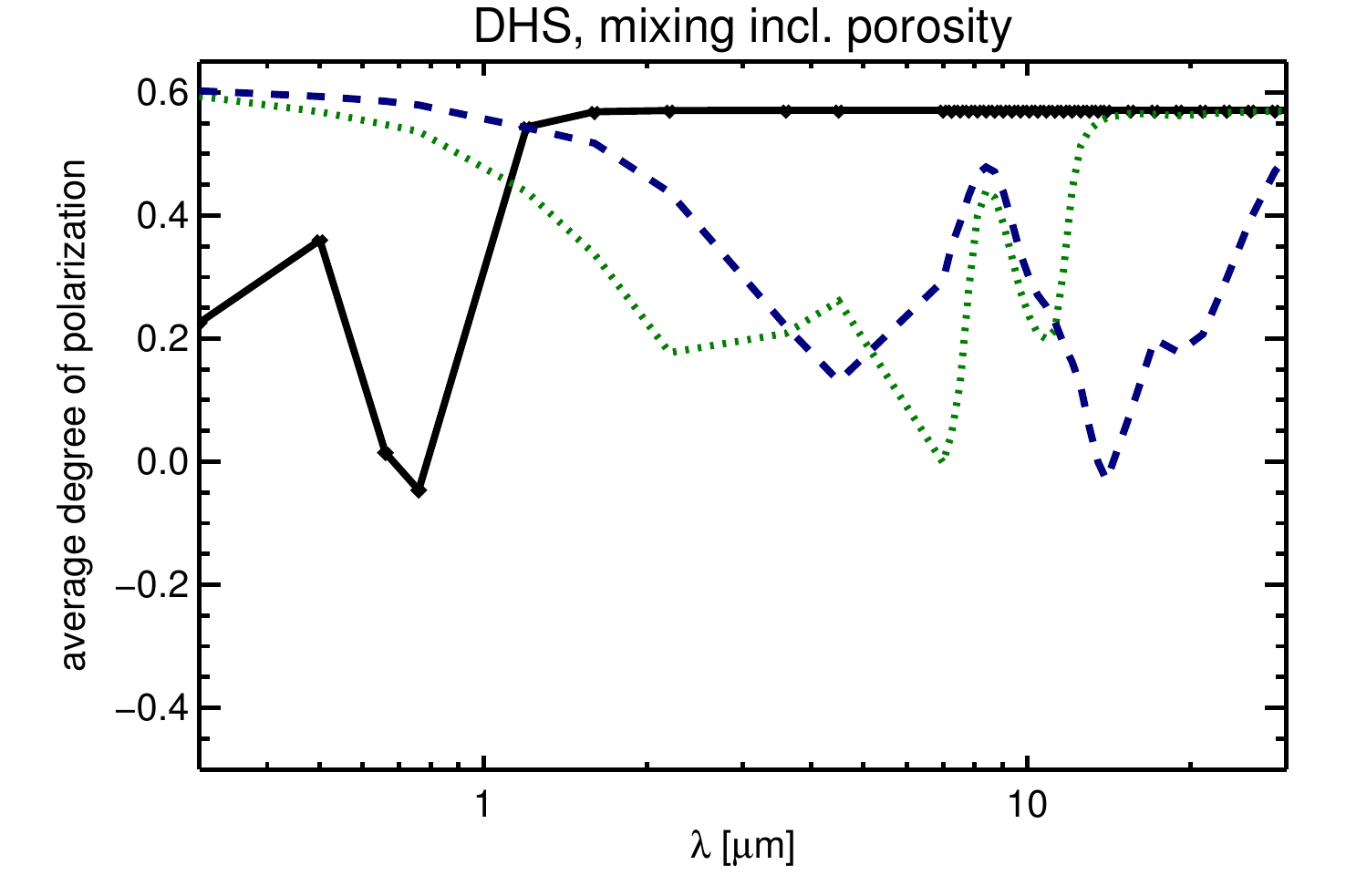}}}
\caption{Degree of linear polarization as a function of scattering angle (upper left panel). The other panels show the scattering-angle-averaged degree of polarization as a function of wavelength for the different particle models (see Eq.~\ref{eq:pav}). Shown are the curves for $r_V=0.2, 2.0,$ and $4.0\,\mu$m.}
\label{fig:polarization}
\end{figure*}

\subsubsection{Polarization properties}
\label{sec:results:pol}

It is often assumed that the degree of polarization is a good indicator of particle size. Particles much smaller than the wavelength of radiation reach a degree of polarization at $90^\circ$ scattering that is 100\%. As the particle size increases, the polarization first drops. However, for very large, smooth, and convex particles the polarization increases again as a result of scattering at the Brewster angle. For a non-absorbing sphere this causes 100\% polarized light at a scattering angle for which $\tan(\pi-\theta/2)=n$, with $n$ the refractive index. This effect does not play a role in scattering off a rough or concave surface, therefore the degree of polarization of realistic particles is not expected to increase at the Brewster angle. Indeed, as we show in the upper left panel of Fig.~\ref{fig:polarization}, the degree of polarization is constant with size. This effect can also be observed for other wavelengths. To examine the behavior of the degree of polarization with wavelength, we define the average degree of polarization as
\begin{equation}
\label{eq:pav}
P_\mathrm{av}=\frac{1}{4\pi}\int_{4\pi} -F_{12}/F_{11}~d\Omega.
\end{equation}
For pure Rayleigh scattering $P_\mathrm{av}=\pi/2-1\approx0.57$. In Fig.~\ref{fig:polarization} we plot the average degree of polarization as a function of wavelength for the aggregates and the simplified particle shapes. It is striking to see that the behavior as a function of wavelength is very simple and smooth for the aggregate particles, while for all simplified approaches the behavior is much more chaotic as a function of wavelength. For the large, smooth particles with mixing, the effect of the Brewster angle becomes very clear as well. This causes a very high polarization for large particles at short wavelengths, which is not seen for the aggregate particles.  This is not seen either when we add the opacities of the single materials because here we add scattering matrices of particles that do not have the same Brewster angle, and particles that do not show strong polarization around the Brewster angle because of strong absorption from the carbon and iron sulfide particles. These effects are washed out to a certain degree in the mixing method. It can be argued that the DHS method comes closest to the aggregate polarization properties with added opacities, but this method does not catch the essentially simple behavior of the aggregate particles either.

\section{Large aggregates and comparison to other approaches}
\label{sec:DIANA}

We have shown how to best approximate the optical properties of dust aggregates in protoplanetary disks. The best match with the exact aggregate computations was reached with the DHS model with effective medium theory and a porosity of 25\%. The next step is to analyze the influence of this different approach on what is currently considered to be the standard. Standards are crucial in modeling. especially for the optical properties of dusty environments. Since the optical properties are the basis of any interpretation, comparing the results from different studies becomes extremely difficult when the underlying assumptions on the optical properties are very different.

In many current modeling approaches, the method most commonly applied is the Mie theory, homogeneous spheres, with a composition characterized by the so-called astronomical silicate \citep{1984ApJ...285...89D}. We compare our new approach to this and show the differences.

Dust particles in protoplanetary disks are expected to grow significantly beyond the limiting size we used so far. With the approach using DHS and effective medium theory we can easily extend the size range to much larger sizes. For the dust size distribution we take a simple power law given by
\begin{equation}
\label{eq:sizedis}
n(r)dr\propto \begin{cases} 
r^{-p}, &       r_\mathrm{min} < r < r_\mathrm{max}, \\
0,              &       \mathrm{elsewhere.}
\end{cases}
\end{equation}
For the computations in this paper we fixed the minimum and maximum size of the distribution to $r_\mathrm{min}=0.05\,\mu$m, $r_\mathrm{max}=3000\,\mu$m.

\begin{figure}[!tp]
\centerline{\resizebox{\hsize}{!}{\includegraphics{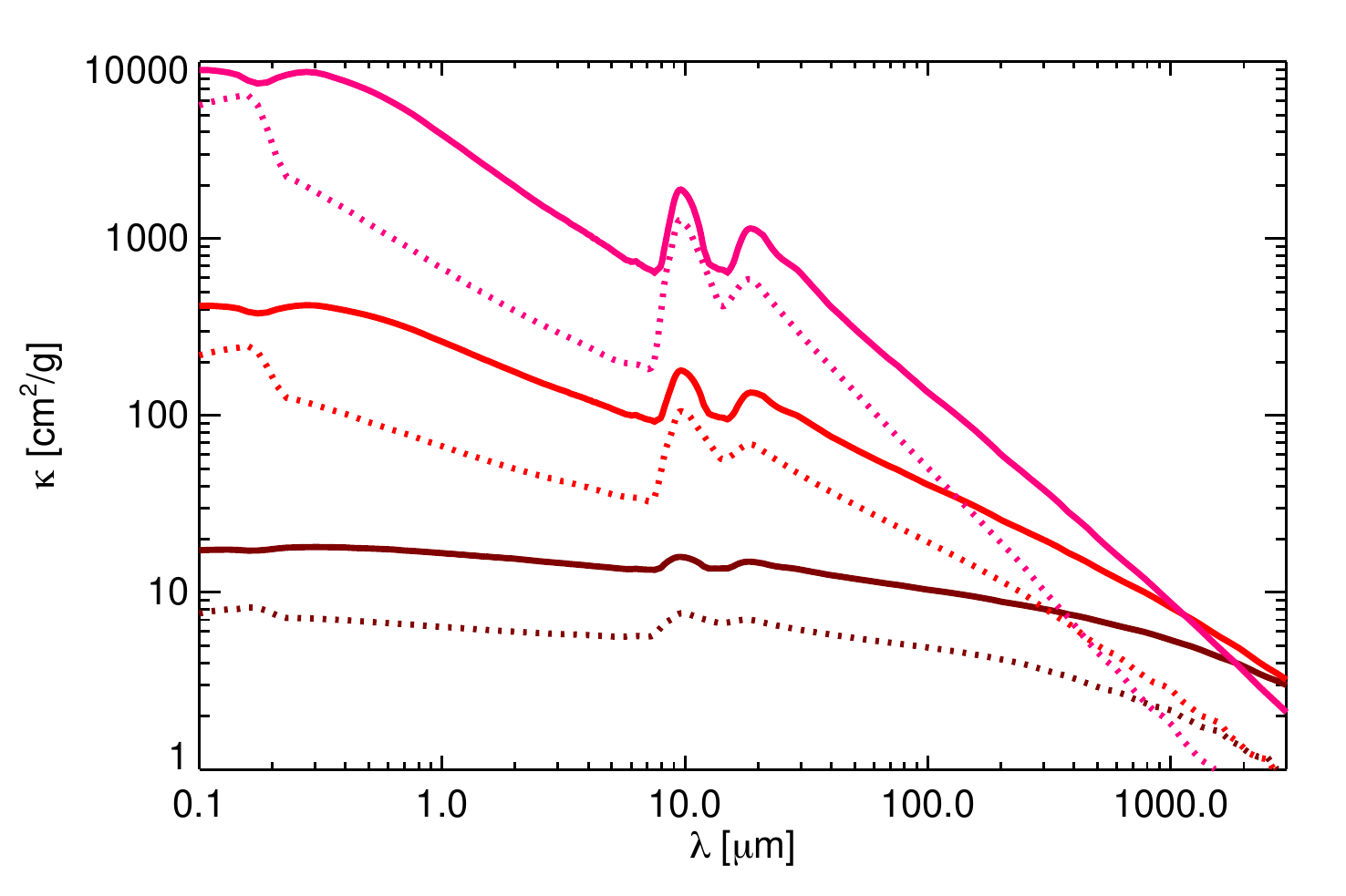}}}
\caption{Size-distribution-averaged mass absorption cross section computed with the DHS method with the effective medium theory (solid lines) compared to those computed for spherical particles composed of astronomical silicate (dotted lines). The different colors correspond to different slopes (i.e., values of $p$) of the size distribution: $p=4$ (pink), $p=3.5$ (red), and $p=3$ (brown).}
\label{fig:sizes}
\end{figure}

In Fig.~\ref{fig:sizes} we show the mass absorption cross sections as a function of wavelength for different values of the slope of the size distribution, $p$. We compare the DHS method with
the effective medium theory and the material mixture used throughout the paper with the curves computed using spherical astronomical silicate particles with the same size distribution. There are a few notable differences. First, the absorption cross section of the entire curve is much lower for the astronomical silicate particles. Second, the slope in the millimeter part of the spectrum is different. How different this slope is depends strongly on the exact choice of the optical properties of the carbon and iron sulfide particles and on their abundance in the mix. Finally, the amorphous silicate feature in the astronomical silicate curves is stronger. All these differences will lead to a significantly different interpretation of observations when using the simplified spherical astronomical silicate grains as compared to the DHS grains with a realistic mixture of laboratory measured materials. The absolute offset and the strength of the silicate feature are most likely caused by the absence of a material like carbon or iron sulfide in the astronomical silicate approximation. The slope of the millimeter opacity is only significantly adjusted with the combination of an adjusted material mixture and particle shape.

\section{Conclusions}
\label{sec:conclusions}

We have computed the optical properties of aggregate particles and compared them to several simplified approaches for the computation of optical properties. The aim was to determine which approximate numerical method can be used to accurately derive the optical properties of particles that are expected to be aggregates, such
as the dust particles in protoplanetary disks or comets. We discussed the general characteristics of aggregate particles and what distinguishes them from compact or smooth particles.

We conclude the following:
\begin{itemize}
\item[$\bullet$] The cross sections and the overall shape of the scattering phase function of aggregate particles are well represented using a combination of effective medium theory, porosity, and the DHS shape distribution.  In this way the internal complexity of the aggregate (through the effective medium theory), the fluffy structure (through the porosity), and the irregular shape of the constituents and the aggregate (through the DHS shape distribution) are captured.
\item[$\bullet$] For an aggregate with a relatively open structure we can distinguish between two different sizes: the size of the aggregate as a whole and the size of the constituents. Different scattering properties couple preferentially to one of these two sizes:
\begin{itemize}
\item The asymmetry of the scattering phase function is almost solely influenced by the size of the aggregate as a whole and is therefore a good indicator of aggregate size.
\item The degree of polarization is not significantly influenced by the size of the aggregate and is expected to be determined by the size of the constituent particles.
\end{itemize}
The particles in our simulation are still relatively compact for computational efficiency. It is expected that the differences between the two different sizes present in the aggregate are more pronounced when the aggregate structure is more fluffy.
\item[$\bullet$] Large aggregate particles display a phase function with a strong forward-scattering peak and mildly backward-scattering part of the phase function. This backward-scattering part of the phase function is not found for smooth particles. In geometries where the forward-scattering peak is not visible (for example in mildly inclined disks), the phase function can appear to be purely backward scattering. This effect cannot be modeled using smooth particles.
\item[$\bullet$] The polarization properties of aggregates are, as already mentioned, dominated by the polarization properties of the aggregate constituents. Therefore, when modeling aggregate polarization properties, the shape of the constituents is an essential parameter. Studies that rely on spherical constituents need to be interpreted with caution.
\item[$\bullet$] Comparing the computational recipe we devised to obtain the opacities of aggregate particles to the frequently employed spherical astronomical silicate particles, we conclude that the astronomical silicate spheres have
\begin{itemize}
\item a lower opacity at all wavelengths,
\item a steeper slope in the millimeter, and
\item a stronger 10\,$\mu$m silicate feature.
\end{itemize}
\end{itemize}

\begin{acknowledgements}
We gratefully acknowledge funding from the EU FP7-2011 under Grant Agreement No 284405.
Ch.R. acknowledges funding by the Austrian Science Fund (FWF) under
project number P24790.
The computational results presented have been
achieved in part using the Vienna HPC astro-cluster of the Dept. of
Astrophysics.
\end{acknowledgements}

\bibliographystyle{aa}
\bibliography{biblio}

\end{document}